\documentclass[11pt]{article}
\usepackage{lineno}
\usepackage{graphicx}
\usepackage{subfigure}
\usepackage{color}
\usepackage{amssymb}

\begin{document}

\title{Closed Contour Fractal Dimension Estimation by the Fourier Transform}

\author{
Jo\~{a}o B. Florindo, Odemir M. Bruno\\ 
Instituto de F\'{i}sica de S\~{a}o Carlos (IFSC)\\ Universidade de S\~{a}o Paulo\\ Av.  Trabalhador S\~{a}o Carlense, 400 -
13560-970\\ S\~{a}o Carlos, SP, Brasil\\ 
jbflorindo@ursa.ifsc.usp.br / bruno@ifsc.usp.br
}

\maketitle

\begin{abstract} 
This work proposes a novel technique for the numerical calculus of the fractal dimension of fractal objects which can be represented as a closed contour. The proposed method maps the fractal contour onto a complex signal and calculates its fractal dimension using the Fourier transform. The Fourier power spectrum is obtained and an exponential relation is verified between the power and the frequency. From the parameter (exponent) of the relation, it is obtained the fractal dimension. The method is compared to other classical fractal dimension estimation methods in the literature, e. g., Bouligand-Minkowski, box-couting and classical Fourier. The comparison is achieved by the calculus of the fractal dimension of fractal contours whose dimensions are well-known analytically. The results showed the high precision and robustness of the proposed technique.
\end{abstract}

\section{Introduction}
In the last years, the fractal dimension has become an important tool for characterize objects, forms and surfaces in some areas of the science. The examples of fractal dimension applications are in a wide range and can be found in areas so distinct as medicine \cite{RGD09}, texture analysis \cite{DH01}, geology \cite{BGB09}, botany \cite{PPFVOB05,BPFC08}, materials engineering \cite{CDHLAB03}, electronic \cite{DMJ08}, physics \cite{SMS10}, histology \cite{LFKIKPTI05}, soil analysis \cite{O99}, plant diseases \cite{QJDPM09}, polymer analysis \cite{CCWH10}, among many others. 

The importance of the fractal dimension and consequently its large number of applications can be explained by the fact that this measure gauges, at different scales, how the fractal fills the space in which it is immersed. Thus, the fractal dimension characterizes the fractal object with a number which is not dependent from the observation scale. As a consequence of this multiscale characteristic, this measure can be still used to detect the level of self-similarity in the analyzed fractal object, that is, it shows how much the object is composed by similar copies of itself at different scales, an intrinsic characteristic of fractals. The self-similarity property is also widely observable in natural objects \cite{M68} as the leaves of a tree or the stream of a river and this fact strongly encourages the researchers in using the fractal dimension to characterize objects from the real world, as in the applications previously cited.

The fractal dimension concept was originally defined in \cite{M68} as being the Hausdorff dimension \cite{F86} of the fractal object. This concept of dimension is an analytical measure whose calculus uses to be complex and even impossible in many cases. Indeed, as the Hausdorff dimension is calculated by a covering process using infinitesimal sets, it is necessary for this calculus to know a well defined mathematical rule used in the formation of the fractal. Such need is not satisfied for example in some Lindenmayer fractals as the three-dimensional weeds \cite{PH89}. As a consequence, along the years, a lot of numerical methods were developed in order to estimate the fractal dimension in these cases. These methods became still more important with the representation of fractal objects in digital images. In the digital representation, the fractals are shown in a discrete space (image grid) which derails the Hausdorff dimension calculus once the fractal is handled as a disconnected set of points, compromising seriously the resultant dimension.

The approximated fractal dimension estimation methods can be split into two approaches. The first are the spatial methods, as box-counting \cite{bb22329}, Bouligand-Minkowski \cite{T95}, mass-radius \cite{LR93}, dividers \cite{SS94}, among others, based on geometric strategies used for measure the self-similarity of the object. The second approach is based on spectral measures, as Fourier \cite{R94} and wavelets \cite{JJ01}. Particularly, fractal dimension estimative methods based on Fourier are interesting once these methods have easy implementation and low computational cost using Fast Fourier Transform (FFT). Another advantage of such techniques is the fact that they are less sensible to geometric transformations such as scale, rotation and translation \cite{R94}\cite{B74}. 

This work proposes a novel method based on Fourier fractal dimension to calculate the fractal dimension of contours. In this technique, the fractal object must be representable by a closed contour and this contour is mapped onto a complex signal. The power law of the power spectrum is used to estimate the fractal dimension using a strategy similar to that described in \cite{C95}. The proposed method is compared in the calculus of the fractal dimension with other fractal dimension methods. Although the work presented here is a theoretical one, and just mathematical fractals was considered, the method can be used to estimate the fractal dimension of signals (in the real or complex domain) and in this way, can be used in a wide range of applications. 

This paper is divided into 7 sections, including this introduction. The following discusses the theory of fractal dimension. The third addresses the Fourier fractal dimension. The forth describes the proposed method. The fifth explains the experiments. The sixth shows the results of the experiments and the last section does the conclusion.

\section{Fractal Dimension}

Fractal objects were studied by mathematicians like Cantor and Peano, since the 19$^{th}$ century, when they were known as monster curves. Such strange structures did not obey the rules of the traditional Euclidean geometry. For instance, in 1904, Helge Von Koch presented a curve composed by an infinite sequence of equalizer triangles and which showed finite area and infinite perimeter. Such condition represented a contradiction to the Euclidean classical rules \cite{BM89}. 

Although these objects brought a certain interest in the principle, it was only with the arising of computers in 1970 decade that the fractal theory assumed a fundamental role. In that years, Mandelbrot \cite{M68} was the first author to formalize the concept of fractal. He defined fractals as organized structures which have infinite complexity, present self-similarity (in some level) and whose Hausdorff dimension exceeds strictly the topological (Euclidean) dimension.

The self-similarity suggests that a worthy measure in the characterization of fractals must present a unique value independent from the observation scale of the object, once we observe the geometric aspect preserved when we analyze the fractal more and more microscopically. The measure satisfying this requirement is the fractal dimension. The Hausdorff dimension \cite{F86} used in the definition of Mandelbrot is an example of such measure which captures the whole range of scales by covering (feeling) the fractal with infinitesimal sets.

For the definition of the Hausdorff dimension, we must initially define the $s$-dimensional Hausdorff measure $H^{s}(F)$ of a subset $F \in \Re^{n}$:
\[
	H^{s}(F) = \lim_{\epsilon \rightarrow 0}{H_{\epsilon}^{s}(F)},
\]
where
\[
	H_{\epsilon}^{s}(F) = inf\left\{ \sum_{i=1}^{\infty}{|U_{i}|^{s}:{U_{i} \mbox{ is an $\epsilon$-cover of F}}} \right\},
\]
in which $||$ denotes the diameter in $\Re^{n}$, that is, $|U| = sup{|x-y|:x,y \in U}$. The Hausdorff dimension properly is obtained through the following expression:
\[
	dim_{H}(F) = \{s\} | \inf \left\{ s:H^{s}(F)=0 \right\} = \sup \left\{ H^{s}(F)=\infty \right\}.
\]

A more practical definition of the Hausdorff fractal dimension can be provided from a generalization of the concept of topological dimension applied to Euclidean objects.

\begin{figure}[!htb]
\centering
\includegraphics[width=0.6\textwidth]{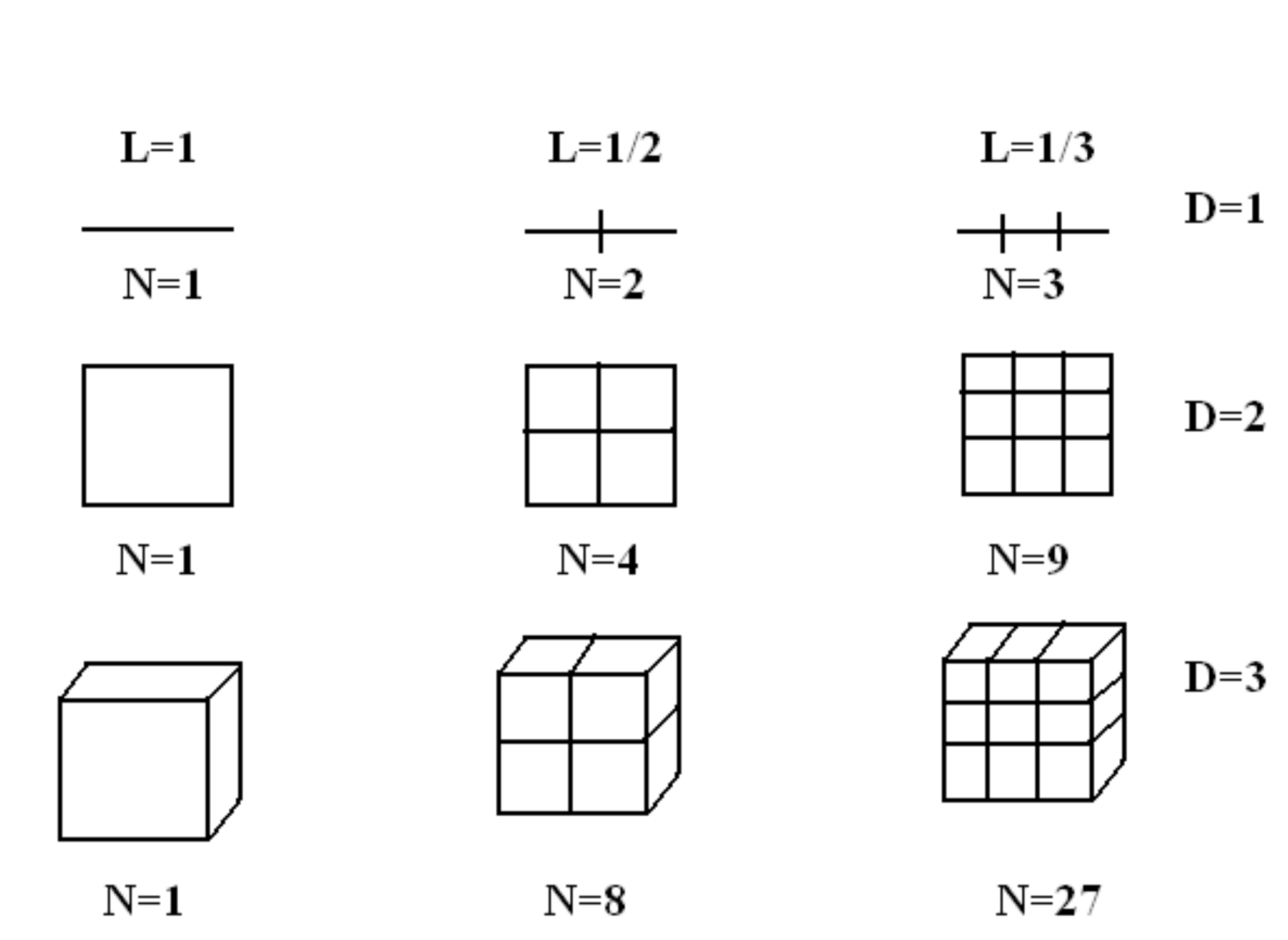}
\caption{The relation between the scale factor L and the number N of objects covering the original structure in the dimension D.}
\label{fig:dim}
\end{figure}

\begin{figure}[!htb]
\centering
\includegraphics[width=0.8\textwidth]{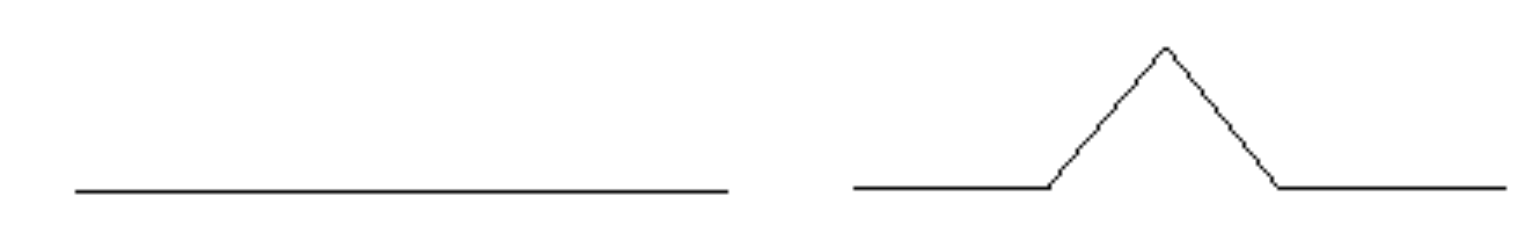}
\caption{The first iteration in the composition of the Koch snowflake curve. 4 segments with one third of the length of the anterior segment replacing it.}
\label{fig:koch}
\end{figure}

In fact, in Euclidean geometry, if we have an object $A$ of dimension $D$ and an object $B$ which is the same object $A$, but with its linear size reduced in all the $D$ directions by $1/l$, the number of objects like $B$ needed to cover (fill) the object $A$ is given by $N(l) = l^{D}$. Notice that this formula is not dependable from the absolute linear size of $A$ and $B$. The Figure \ref{fig:dim} illustrates this fact in Euclidean geometry. In this way, the dimension may be calculated by:
\[
	D = \frac{log(N(l))}{log(\frac{1}{l})}.
\]
Extending to fractal structures, we have the following simplified version of the Hausdorff dimension:
\[
	D = \lim_{\epsilon\rightarrow 0}\frac{log(N(\epsilon))}{log(\frac{1}{\epsilon})},
\]
in which $N(\epsilon)$ is the number of objects with linear size $\epsilon$ needed to cover the whole object \cite{F86}. For example, in the Koch curve, we have at each iteration, the curve covered by 4 segments whose length is 1/3 of that segment in the anterior step. The Figure \ref{fig:koch} illustrates the process. Applying the dimension formula for each iteration $k$, we have:
\[
	D = \lim_{\epsilon\rightarrow 0}\frac{log(N(\epsilon))}{log(\frac{1}{\epsilon})} = \lim_{k\rightarrow\infty}\frac{log(4^k)}{log(\frac{1}{1/3})^k} = \frac{log(4)}{log(3)} \approx 1.26.
\]

The Hausdorff dimension in the most of cases presents a complex mathematical calculus and in many cases this calculus is impossible, like when we do not know the rules of composition of the fractal. In order to fill this gap, along the years, a lot of numerical methods have been developed for the calculus of the fractal dimension in general situations. These methods can be split into two approaches: the spatial and the spectral techniques. The first is based on spatial measures as box-counting \cite{bb22329}, Bouligand-Minkowski \cite{T95}, mass-radius \cite{LR93}, dividers \cite{SS94}, among others. The second approach is based on operations on the frequency domain, as Fourier \cite{R94}, wavelets \cite{JJ01}, etc.

From the spatial methods, we explain briefly the operations of the two most used in the literature, that is, box-counting and Bouligand-Minkowski.

\subsection{Box-counting}

For the definition of the fractal dimensions in $\Re^2$ here presented, we consider the fractal object as a nonempty subset in $\Re^2$. The essential definition of the box-counting dimension $dim_{B}C$ of $C$ is therefore given by:
\[
	dim_{B}C = \lim_{\delta \rightarrow 0}\frac{N_{\delta}(C)}{-log \delta},
\]
where $N_{\delta}(C)$ is the minimum number of sets with diameters at least $\delta$ needed to cover $C$. With the aim of simplifying the numerical calculus, we can consider $N_{\delta}(C)$ as the number of $\delta$-mesh boxes $b$ which intersect $C$, where $b$ is a $\delta$-coordinate of $\Re^2$ given by:
\[
	b = [b_{1}\delta,(b_{1}+1)\delta \times b_{2}\delta,(b_{2}+1)\delta],
\]
where $b_{1}$ and $b_{2}$ are integers.

\begin{figure}[!htpb]
\centering
\includegraphics[width=0.8\textwidth]{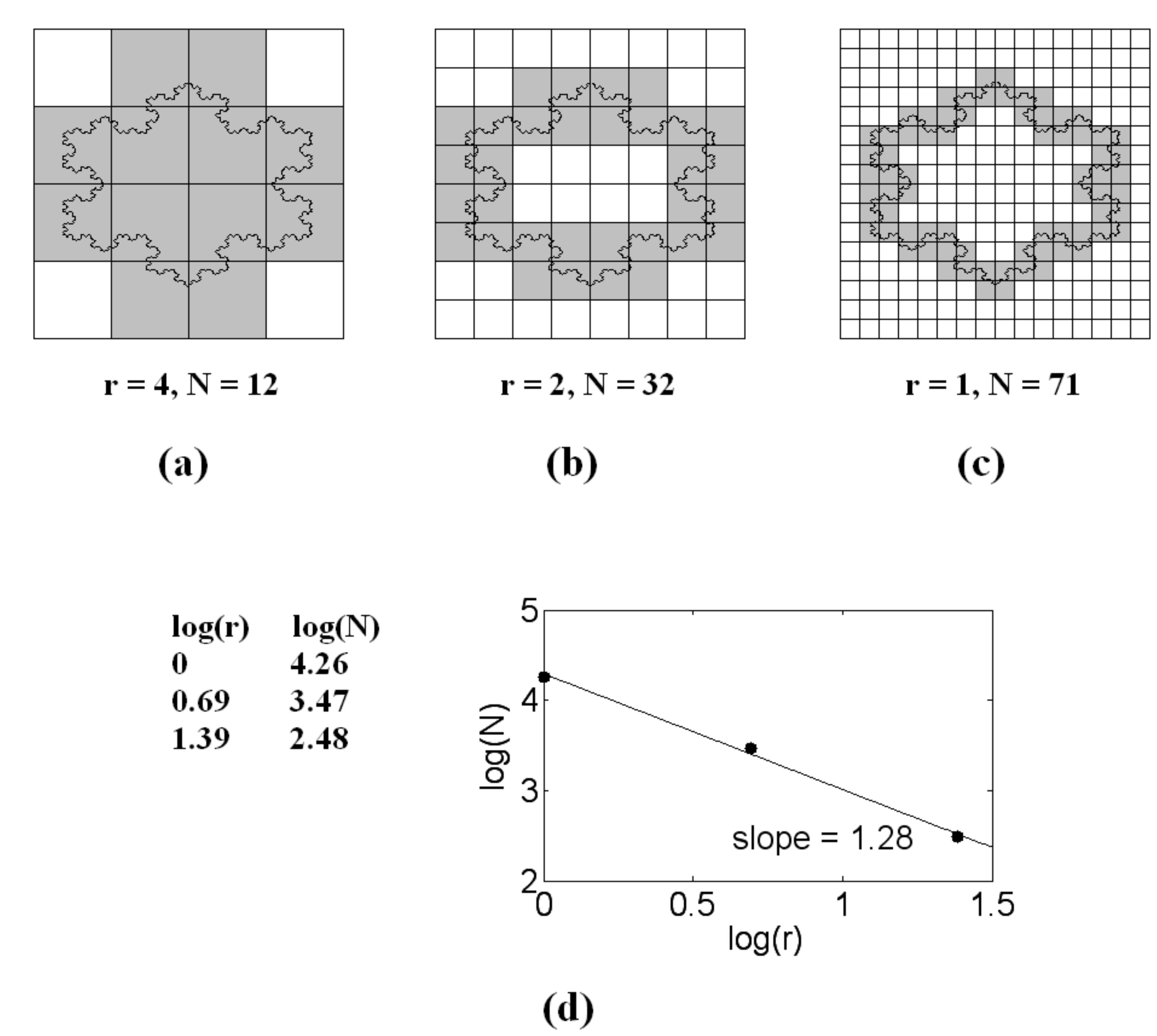}
\caption{The grid used in box-counting method with different size of each square and the squares intersected by the contour shaded in each case. The square size is represented by $r$ and the number of intersected squares by $N$. (a) Grid with ($r = 4$). (c) Grid with $r = 2$. (d) Grid with $r = 1$. (d) Table of $r$ by $N$ and graphical of $log(r) \times log(N)$ fitted by a stright line and its slope.}
\label{fig:box}
\end{figure}

For practical implementations, the box-counting method is based on the division of the space of the object to be analyzed into a grid of squares (boxes). At each iteration of the method it is used a different side length $r$ for the boxes in the grid. The fractal dimension is given by $-\alpha$ where $\alpha$ is the slope of a straight line fitted to the curve of the function $N(r)\times r$ in a log-log scale, where $N(r)$ is the number of boxes of length $r$ which intersect the fractal object. The Figure \ref{fig:box} illustrates the process.

\subsection{Bouligand-Minkowski}

In the case of the Bouligand-Minkowski method, we initially must define the Bouligand-Minkowski measure of $C$ given by:
\[
	meas_{M}(C,S,\tau) = \lim_{r \rightarrow 0}\frac{V(\partial X \oplus rS)}{r^{n-\tau}},
\]
where $S$ is a structuring symmetrical element with radius $r$ and $V$ is the volume of the dilation between $S$ and the boundary $\partial C$ of $C$. The Bouligand-Minkowski dimension depends only upon the structuring element used and it is thus obtained through:
\[
	dim_{M}(C,S) = \inf \left\{ \tau,meas_{M}(C,S,\tau) = 0 \right\}.
\]

For practical purposes, the Bouligand-Minkowski dimension is calculated through neighborhood techniques. Therefore, in cases where we analyze curves in the $\Re^2$ as is the case in this work, each point of the curve is replaced by a disk $S_{\epsilon}$ with diameter $\epsilon$ composing a dilated area for each value of $\epsilon$. The fractal dimension in this process can be calculated by the following expression:
\[
	dim_{M}(C) = \lim_{\epsilon \rightarrow 0}\left( 2-\frac{log A(X \oplus S_{\epsilon})}{log \epsilon} \right).
\]

\begin{figure}
					 \centering
							\includegraphics[width=0.8\textwidth]{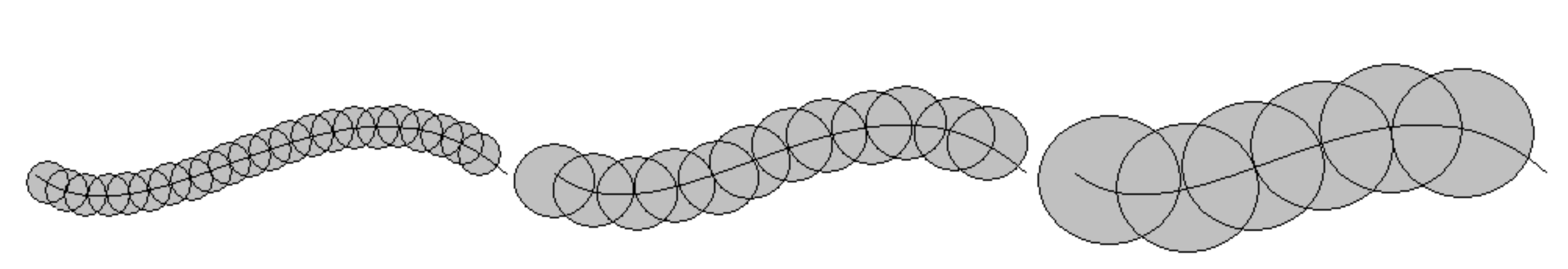}				 
           \caption{Bouligand-Minkowski method. A curve surrounded by circles with different radii composing the dilation area (shaded). The dilation is composed by the union of the circles and inasmuch the radius is increased, the circles begin to interfere in each other. The interference pattern follows the multiscale properties of the fractal allowing the calculus of the fractal dimension.}
           \label{fig:mink}                                  
   \end{figure}

In the computational calculus, the Bouligand-Minkowski method is applied by constructing circles with radius $r$ centered at each point of the fractal. The radius varies at each step of the method and the set of points pertaining to the union of the circles of radius $r$ composes the dilation area $A(r)$. With the aim of optimizing, the dilation area is calculated usually by a method named Euclidean Distance Transform (EDT) \cite{FCTB08}.The fractal dimension is given by $N-\alpha$ where $\alpha$ is the slope of a straight line fitted to the curve of the function $A(r)\times r$ in a log-log scale and $N$ is the topological dimension of the Euclidean space in which the fractal object is immersed, in this case, 2. The Figure \ref{fig:mink} illustrates the method.

Given a brief explains of the most commonly used fractal dimension spatial methods, this work will focus on the frequency approach and the next section describes the Fourier fractal dimension calculus.

\section{Fourier Transform and Fractal Dimension}\label{sec:Fourier}

The use of the Fourier transform as a tool for the calculus of the fractal dimension is related to the definition of fractal transform. For our study, we must consider the Fourier transform of a mass distribution $\mu(u)$, $u \in \Re^{n}$, that is, a measure defined on a bounded subset of $\Re^n$, for which $0 < \mathfrak{m}(\Re^n) < infty$. The Fourier transform can thus be obtained through:
\[
	\mathfrak{T}(\mathfrak{m}(u)) = \int_{\Re^n}{e^{ix \bullet u}d\mathfrak{m}(x)},
\]
where $\bullet$ denotes the scalar product defined over $\Re^n$.

From the Newtonian physics, we obtain the gravitational s-potential $\mathfrak{p}_{s}$, $s \geq 0$, of the mass distribution by:
\[
	\mathfrak{p}_{s}(x) = \int{\frac{1}{|x-y|^{s}}d\mathfrak{m}(y)}
\]
Extending the Physics analogy, the s-energy $\mathfrak{e}_{s}$ is obtained through:
\[
	\mathfrak{e}_{s}(\mathfrak{m}) = \int{\mathfrak{p}_{s}(x)d\mathfrak{m}(x)}.
\]
By applying the Fourier transform to $\mathfrak{p}_{s}$ and using the Parseval's theorem:
\[
	\mathfrak{e}_{s}(\mathfrak{m}) = (2\pi)^{n}c\int{\mathfrak{T}(\mathfrak{p}_{s})(u)\overline{\mathfrak{T}(\mathfrak{m}(u))}du},
\]
where $\overline{x}$ is the complex conjugate of $x$, so that:
\begin{equation}\label{eq:energy}
	\mathfrak{e}_{s}(\mathfrak{m}) = (2\pi)^{n}c\int{|u|^{s-n}|\hat{\mathfrak{m}}(u)|^{2}du}.
\end{equation}

If there is a mass distribution $\mathfrak{m}(u)$ on the set $C \in \Re^2$ for which the expression \ref{eq:energy} is finite for some value(s) of $s$, so the Hausdorff dimension of $C$ has its lower limit in $s$. Particularly, if $|\mathfrak{T}(\mathfrak{m}(u))| \leq b|u|^{-t/2}$, for a constant value $b$, then $\mathfrak{e}_{s}(\mathfrak{m})$ always converges if $s < t$. The greatest $t$ for which there is a mass distribution $\mathfrak{m}$ on $C$ is called the Fourier fractal dimension of $C$.

Russ \cite{R94} extended the use of Fourier transform to calculate the fractal dimension of objects represented as surfaces in a digital image. Russ associated the Newtonian potential to the power spectrum of the discrete Fourier transform of the image. Based on this assumption, \cite{R94}\cite{C95}\cite{QMACG08} shows that there is a simple exponential relation between the power spectrum of the Fourier transform of a gray-level image and the frequency variable in the same transform. Such fact can be expressed by the following formula:
\[
	P(f) \propto f^{2-\beta}
\]
It is still verified that the fractal dimension of the surface represented in that image is directly related to the coefficient $\beta$ by the following expression:
\[
\beta = 2H+2,
\]
where $H$ is the Hurst coefficient. This coefficient is directed related to the fractal dimension $FD$ by:
\[
	FD = 3 - H
\]
In this way, one may easily deduce that the fractal dimension of the surface can be obtained through:
\[
	FD = \frac{\beta+6}{2}.
\]

In practice, the Fast Fourier Transform (FFT) algorithm is applied to the image analyzed. The average power spectrum is taken along the frequency (radially) and the curve of the power spectrum against the frequency (distance from the center of the spectrum) is plotted in a log-log scale. A straight line is fitted to the curve and its slope is taken as the value of $\beta$ used in the formula above.

It is also important to notice that for the best efficiency of this method in the calculus of the fractal dimension, the phase at each point of the transform must be the most random possible \cite{R94}. This fact implies that the values of the phase must be totally de-correlated, condition that may be checked by the calculus of a correlation measure along the Fourier phase.
   
\section{Proposed Method}

This work proposes a numerical method for the calculus of the fractal dimension of fractal objects which may be represented by a closed contour. The contour is thus represented by a parameterized curve. Such strategy is largely used since it eliminates redundancies intrinsic to the binary image representation of the contour and identifies more explicitly the most important elements. The general approach to extract the parametric curve is to choose an arbitrary initial point pertaining to the contour and to walk along this contour following a pre-defined direction. Each point visited in the step $t$ is the coordinate $(x(t),y(t))$ of the parametric curve. More details and methods for extract the parametric function of a contour may be seen in \cite{CC00}.

\begin{figure}
					 \centering
					 \mbox{\subfigure[]{\includegraphics[width=0.5\textwidth]{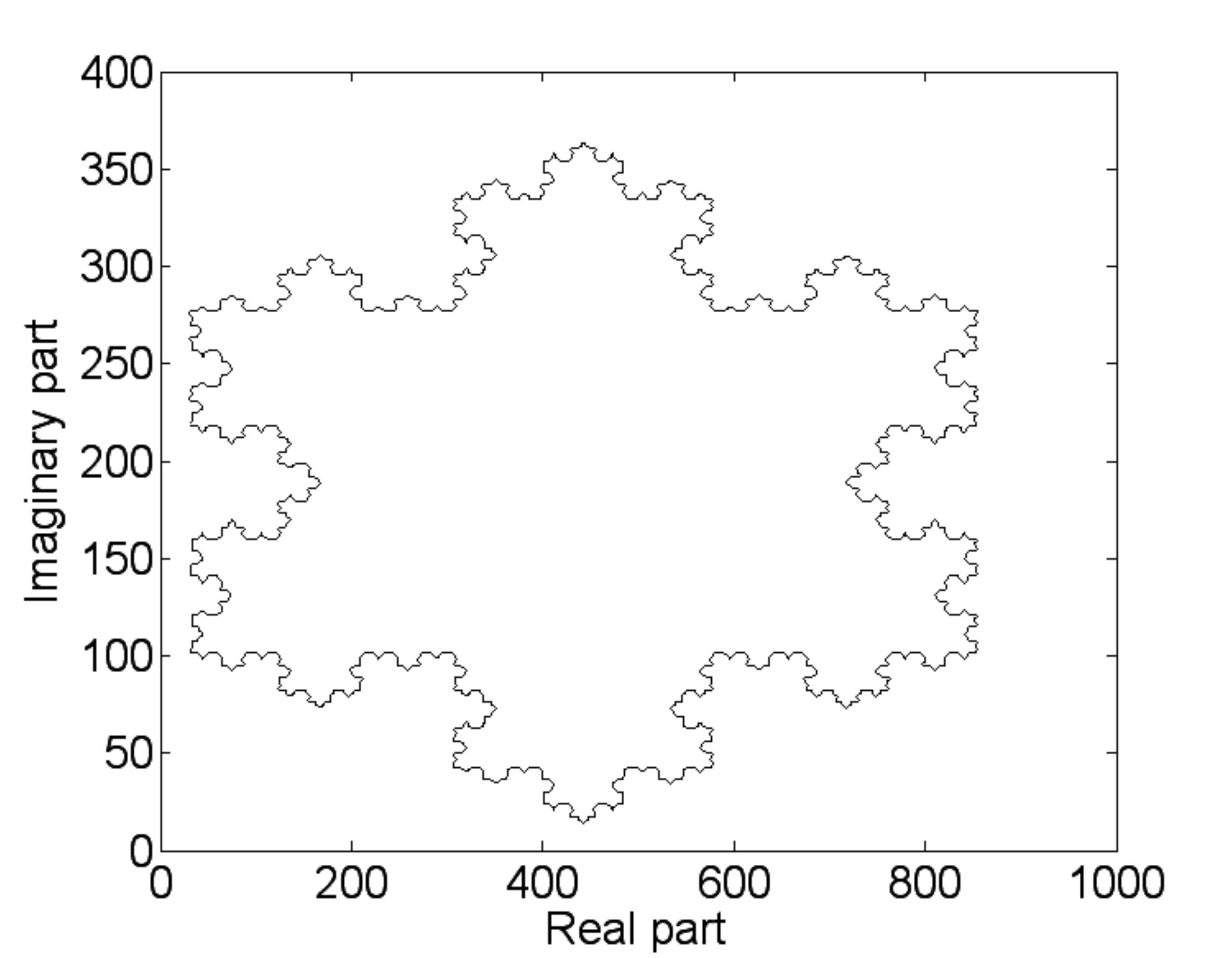}}
								 \subfigure[]{\includegraphics[width=0.5\textwidth]{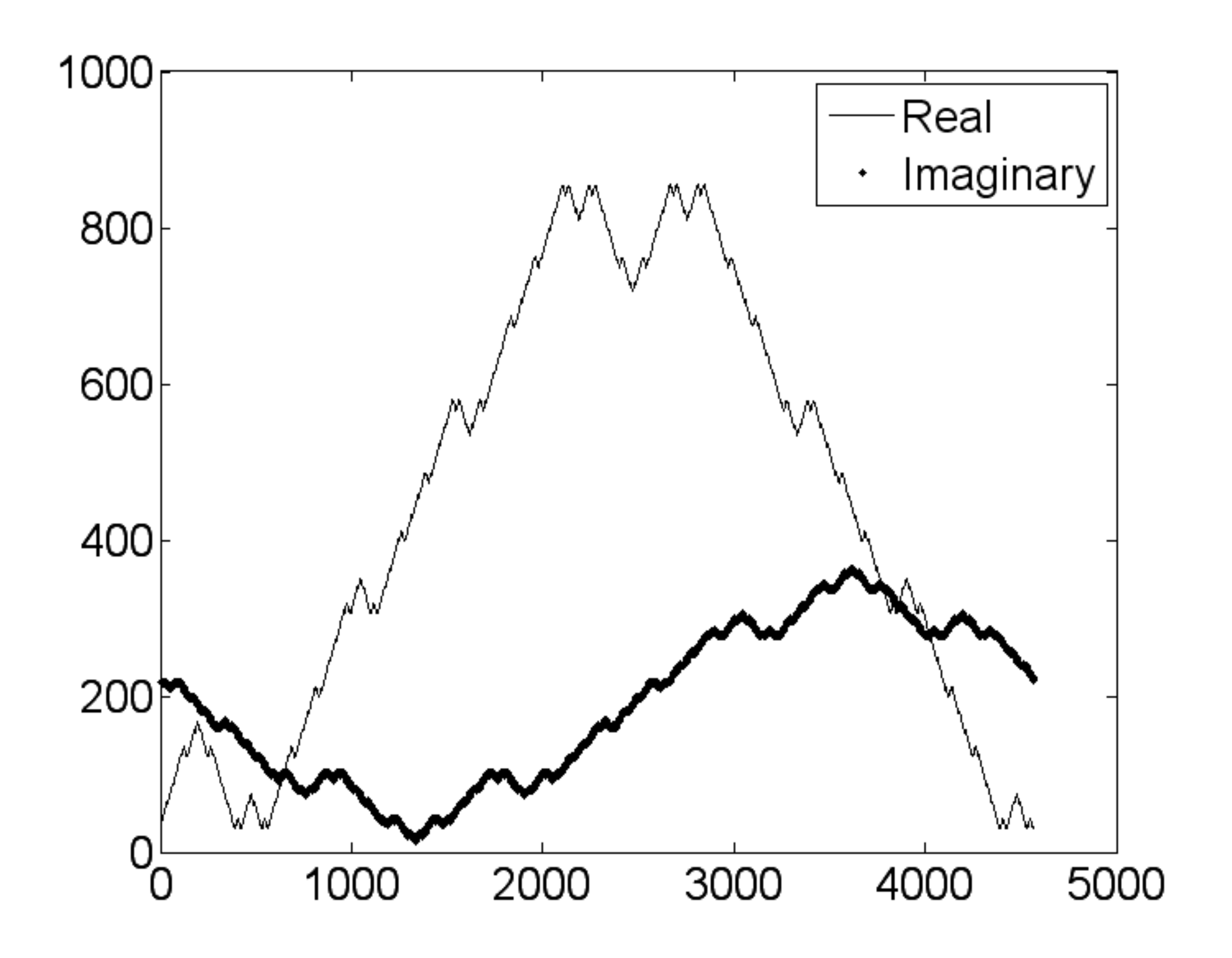}}}					 
           \caption{A contour and its parametric representation. a) The contour of the snowflake fractal. d) Signals obtained from the $x$ and $y$ coordinates of the parametric curve.}
           \label{fig:complexo}                                  
   \end{figure}

Posteriorly, following the order of the obtained contour, each point coordinate $(x(t),y(t))$ is mapped onto a complex-valued signal $C(t)$:
\[
	C(t) = x(t) + iy(t),
\]
where $t$ is the parametric variable of the contour and $i$ is the imaginary number. The Figure \ref{fig:complexo} illustrates the process.

Thus, the Fourier Transform is applied to the signal $C(t)$, obtaining the transformed signal $\tilde{C}(u)$:
\[
	\mathfrak{T}(C(u)) = \int_{-\infty}^{+\infty}{C(t)e^{-2\pi itu}dt},
\]
where $u$ is the frequency variable. As for the classical Fourier dimension described in the Section \ref{sec:Fourier}, the function $C(t)$ can be handled as a mass distribution and like in \cite{R94} the energy is obtained through the spectrum $P$:
\[
	P = |\mathfrak{T}(C(u))|^{2},
\]
where $|.|$ denotes the conventional norm in $\Re^1$. As a consequence, the power spectrum becomes a powerful tool for the calculus of the fractal dimension of the object represented by the signal. The experiments verified that the generalization of Russ also holds in this case and a linear relation is observed between $P$ and $u$, that is:
\[
	P \propto u^{D},
\]
where $D$ is the fractal dimension. From the above expression, we can obtain the following relation for the dimension:
\[
	D \propto log_{u}P.
\]
Applying the well-known basis change property and replacing the proportionality by using constants, we obtain:
\[
	D = k_{1}\frac{log(P)}{log(u)} + k_{2}
\]
where $k_{1}$ and $k_{2}$ are real constants which do not affect the exponential relation. For a precise numerical calculus of $D$ we must make use of intervals with the smallest possible length and calculate:
\[
	D = k_{1}\lim_{h\rightarrow 0}\frac{log(P+h)-log(P)}{log(u+h)-log(u)} + k_{2}.
\]
Disregarded the constants, the above expression represents the derivative of $log(P)$ relative to $log(u)$. And so we deduce that the fractal dimension can be simply obtained from the calculus of this derivative. For the calculus over the discrete signal $C(t)$ the derivative is calculated in the practice by the slope of a straight line fitted to the curve of the graphical of $log(P(u))\times log(u)$.

From the experiments, we obtained the following values for $k_{1}$ and $k_{2}$:
\[
	k_{1} = -\frac{3}{4}; k_{2} = \frac{1}{4}.
\]

\begin{figure}[!htpb]
\centering
\includegraphics[width=0.9\textwidth]{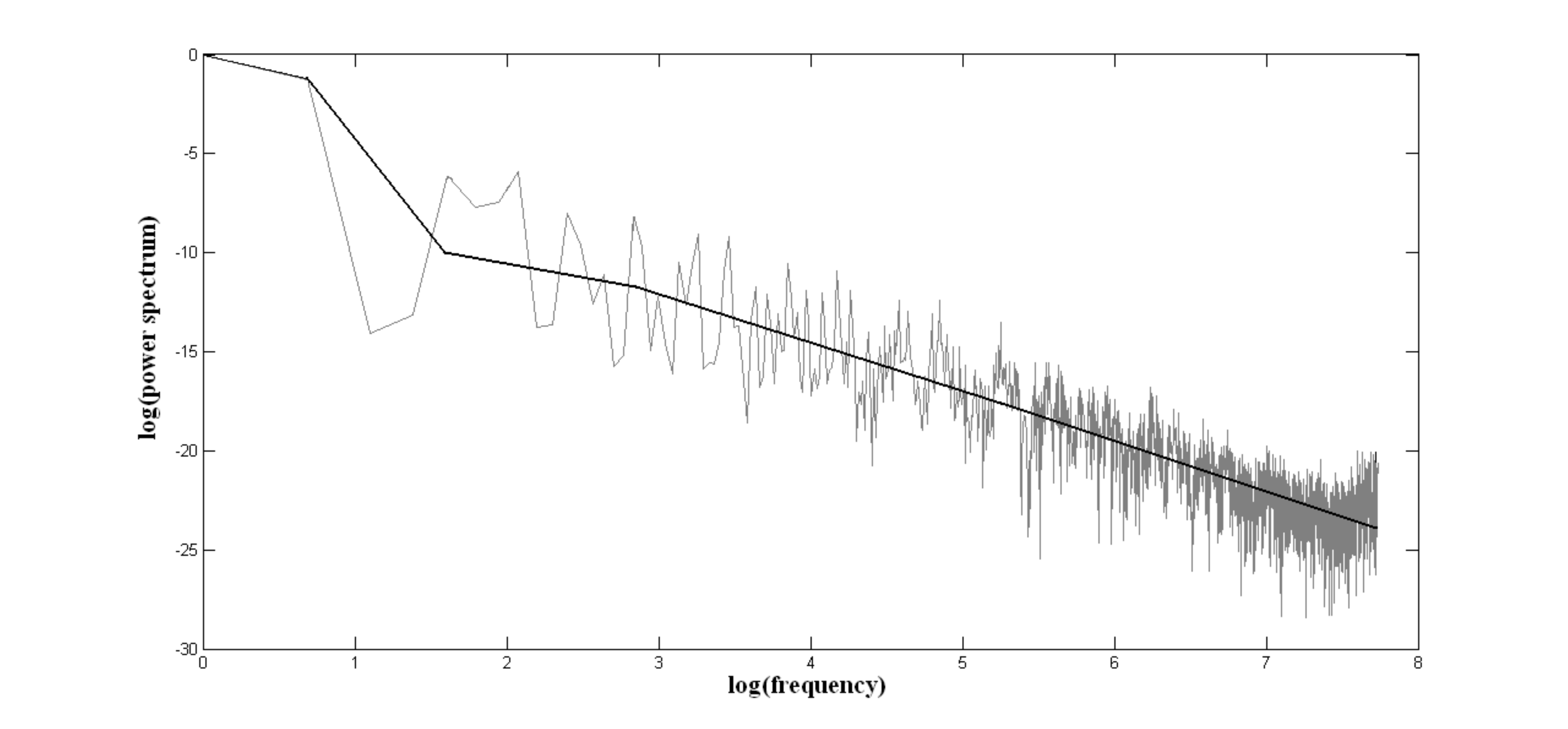}
\caption{The curve of power spectrum$\times$frequency and the slope varying depending on the region of the graph to be considered.}
\label{fig:str}
\end{figure}

We also notice in the experiments a more significant variation of this slope depending on the region taken into account, as one can see in the Figure \ref{fig:str}.

\begin{figure}[!htpb]
\centering
\includegraphics[width=0.9\textwidth]{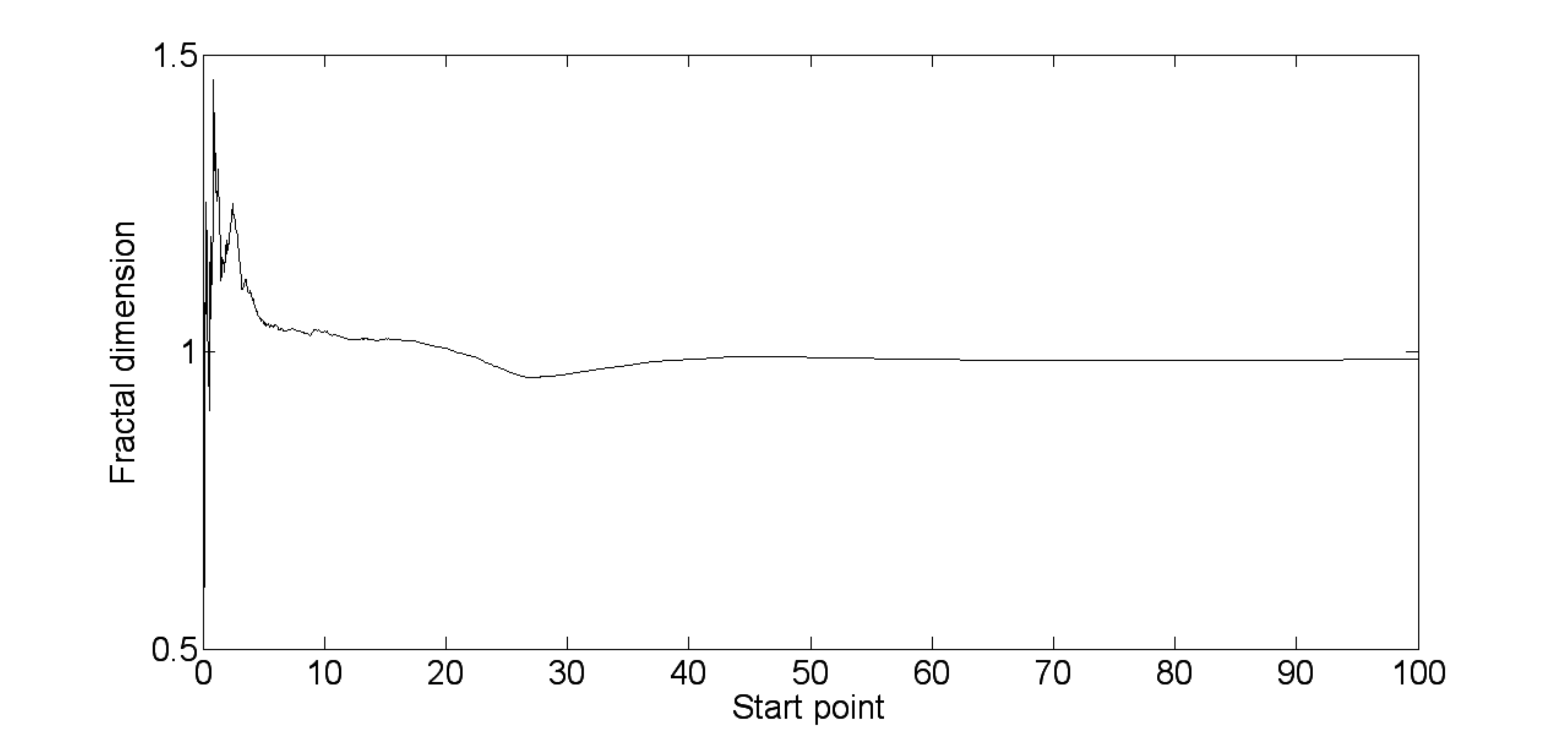}
\caption{A graph representing the slope of the log-log power spectrum curve in function of the starting point considered in the graph.}
\label{fig:graph}
\end{figure}

In the particular case of the fractal objects analyzed in this work, the experiments demonstrated that the slope has a great variation in the first points of the curve and after a certain number of points, the slope tends to remain constant. The Figure \ref{fig:graph} illustrates such behavior by depicting the slope as a function of the first point taken into consideration for the linear fitting.

\begin{figure}
					 \centering
           \mbox{\subfigure[]{\includegraphics[width=0.3\textwidth]{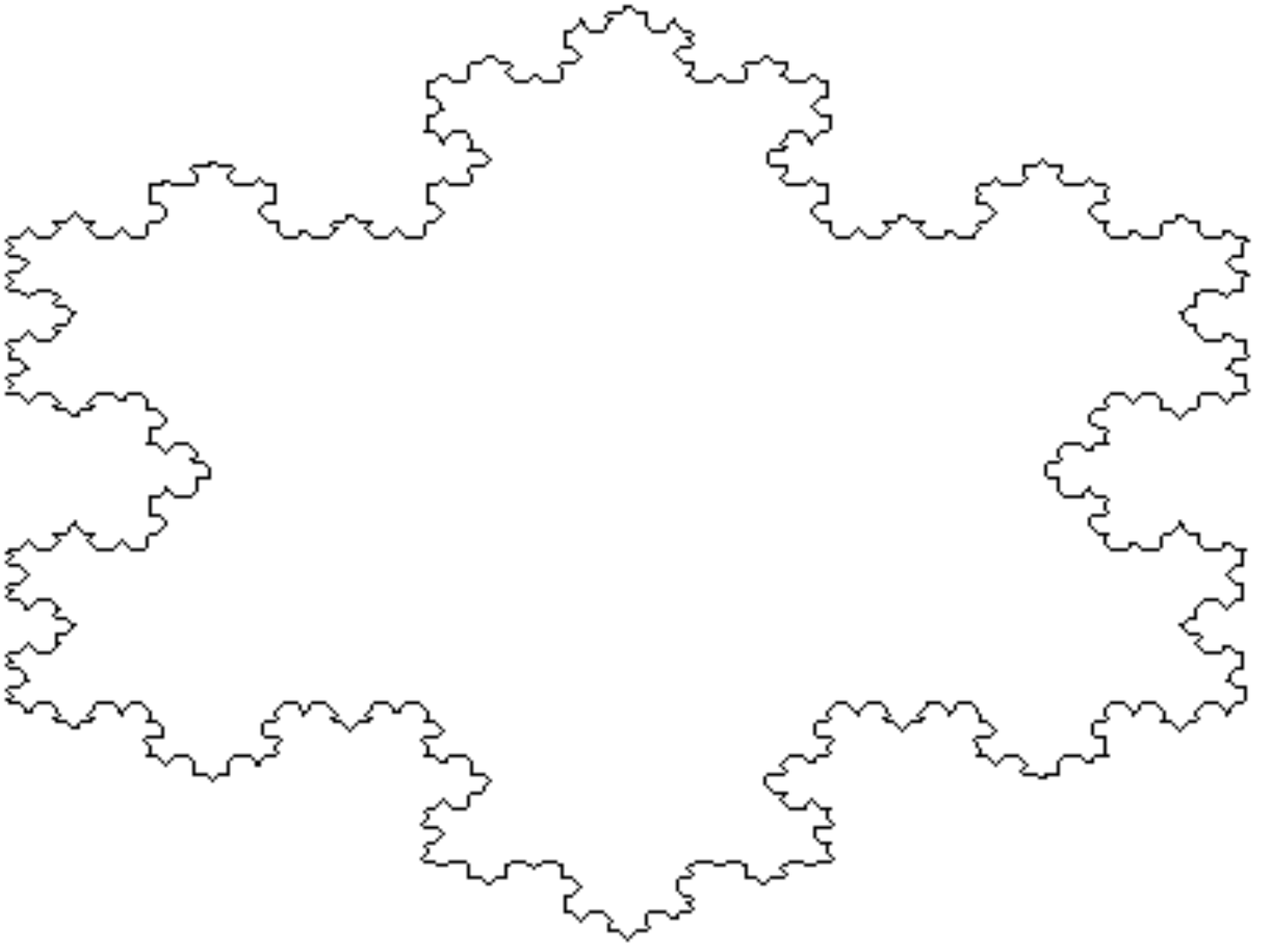}}
           			 \subfigure[]{\includegraphics[width=0.3\textwidth]{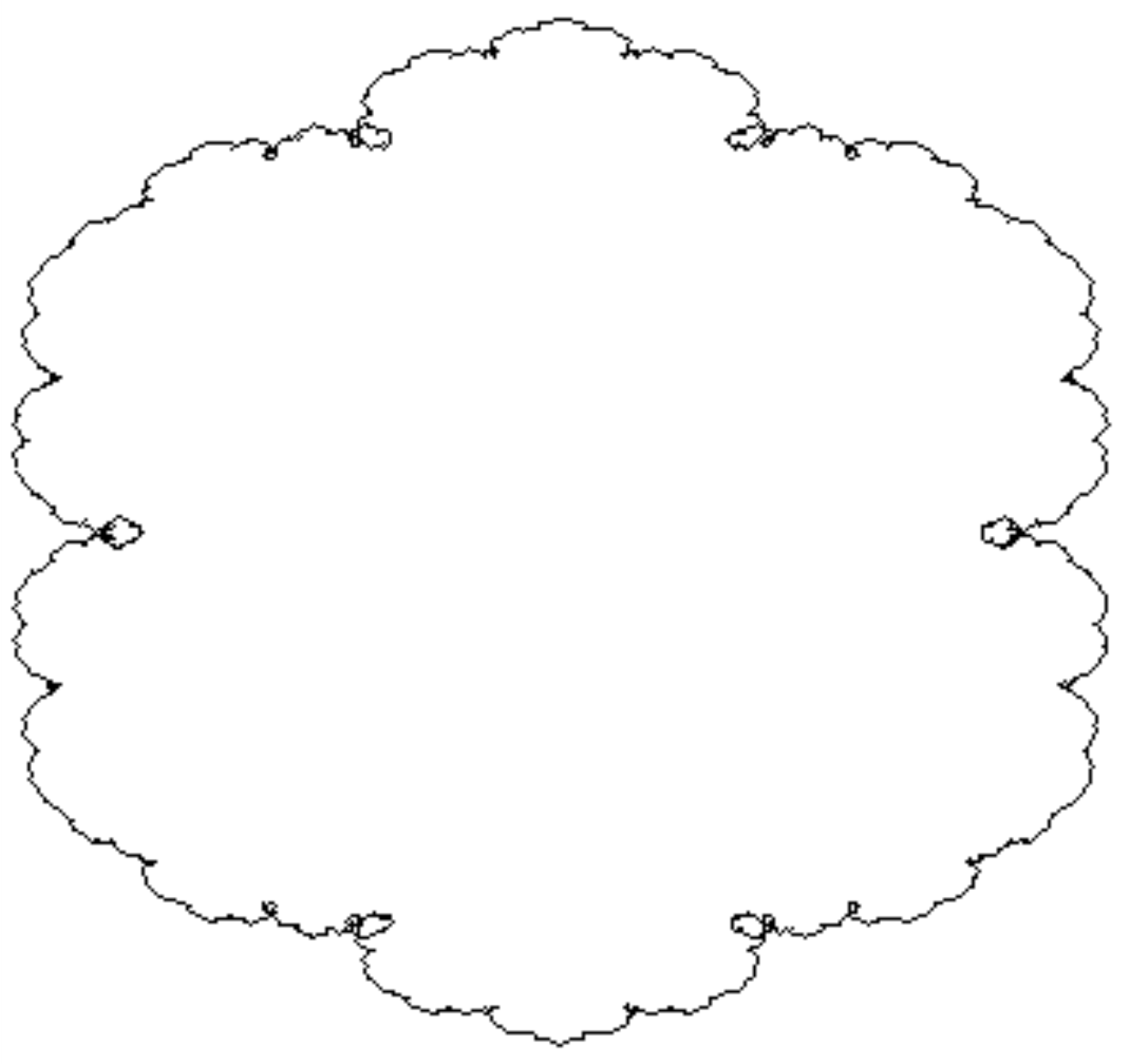}}
           			 \subfigure[]{\includegraphics[width=0.3\textwidth]{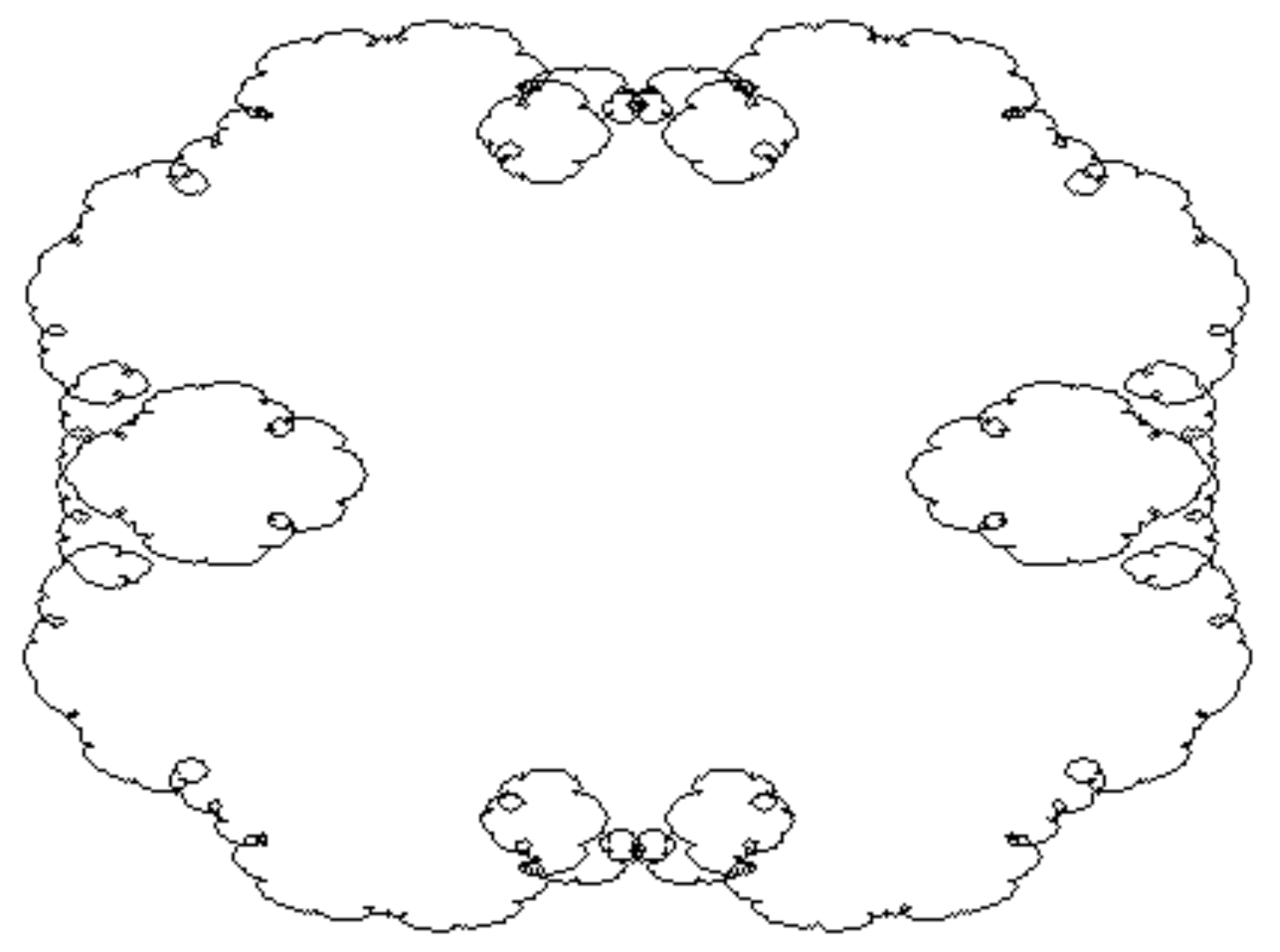}}}									 					
           \caption{Fractal object reconstructed after Fourier transform. a) Original fractal. b) Reconstruction using the first terms. c) Reconstruction using the last terms.}
           \label{fig:bands}                                  
   \end{figure}

Such fact was previsible since the fractal objects contain significant information in its more microscopic (detailed) scales. This interesting property isnot observable in an Euclidean object but is intrinsic to fractals and is justifiable by the complexity inherent to fractals. The Fourier transform preserves the global aspect of the object in the first elements and the details in the last elements. The Figure \ref{fig:bands} illustrates this fact by showing a fractal object (a), its reconstruction using the first elements of the Fourier transform (b) and the using last elements (c). We observe that the first elements preserved the global shape of the fractal while the last ones highlighted the details as the recursive entrances in the fractal shape. In this way, it is convenient the discard of the firt terms in the Fourier transform.
   
For the application developed here, the first 10\% of the points in the transformed curve were discarded, with the aim of capturing the maximum possible of details present in a fractal object. Such fact turns the method here exposed very sensible to imperfections in the structure of the fractal measured.

\section{Experiments}

For the validation of the proposed technique, this work calculates the fractal dimension of fractal objects whose Hausdorff dimension is well-known in the literature.

\begin{figure}
					 \centering
           \mbox{\subfigure[]{\includegraphics[width=0.24\textwidth]{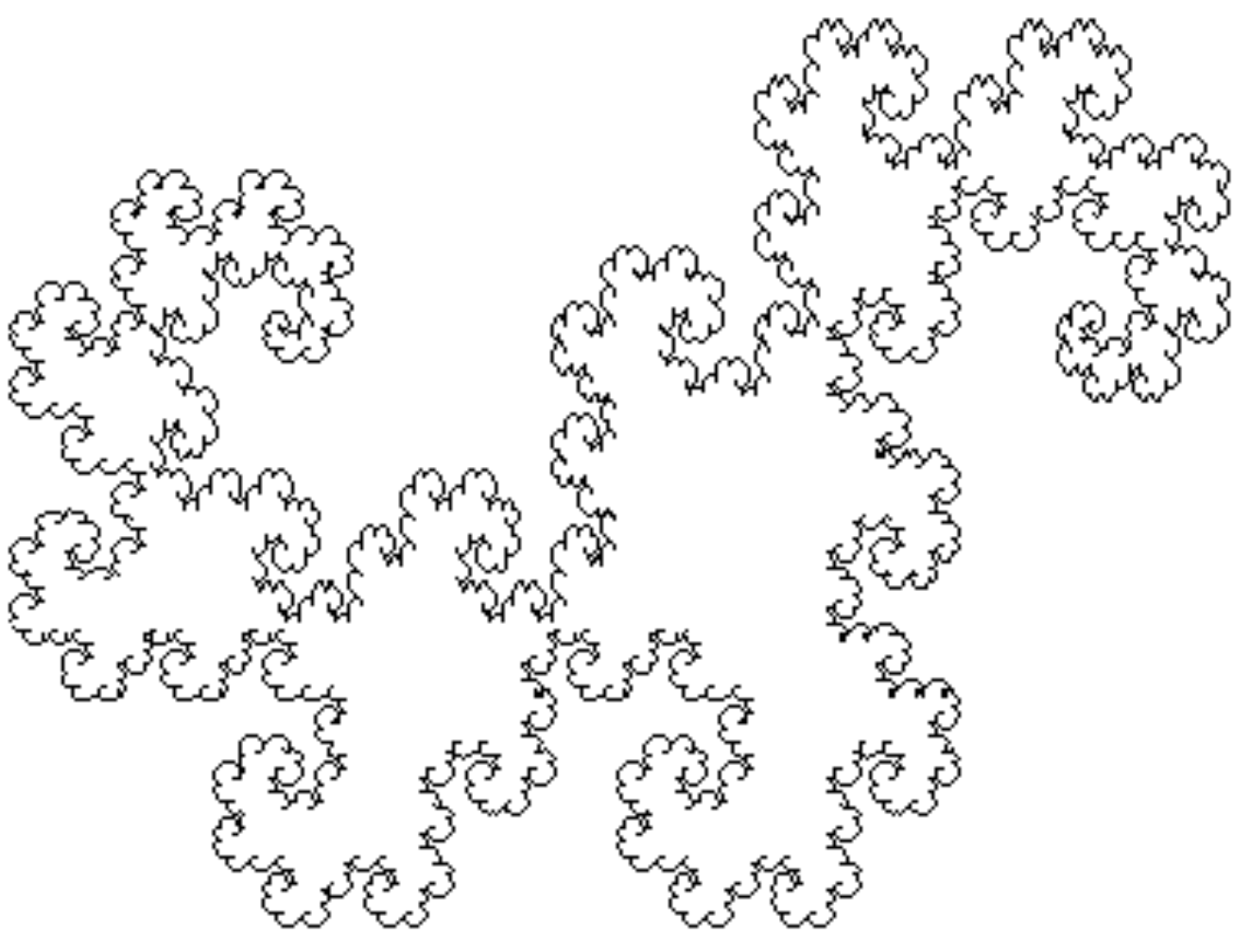}}
								 \subfigure[]{\includegraphics[width=0.24\textwidth]{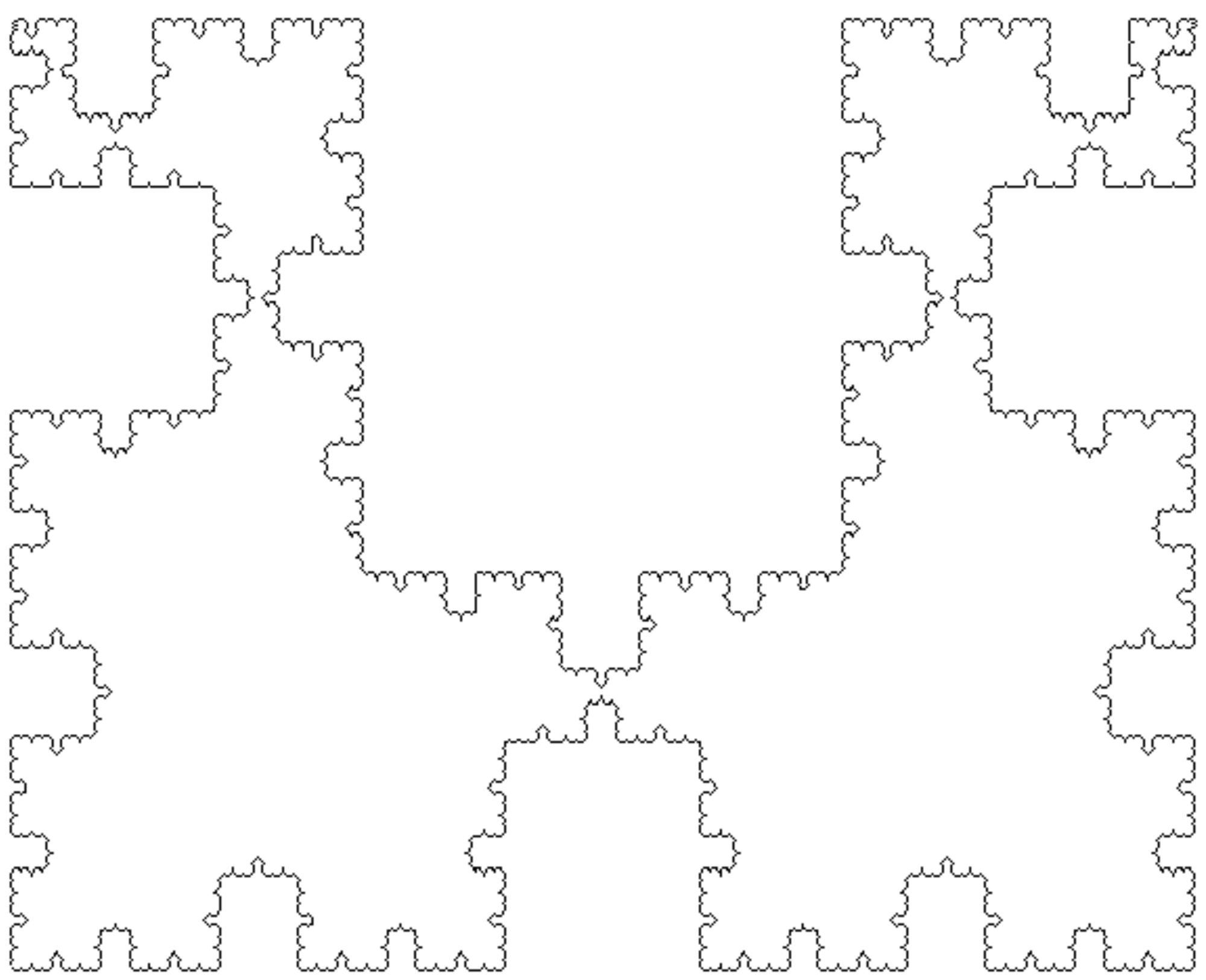}}           
           			 \subfigure[]{\includegraphics[width=0.24\textwidth]{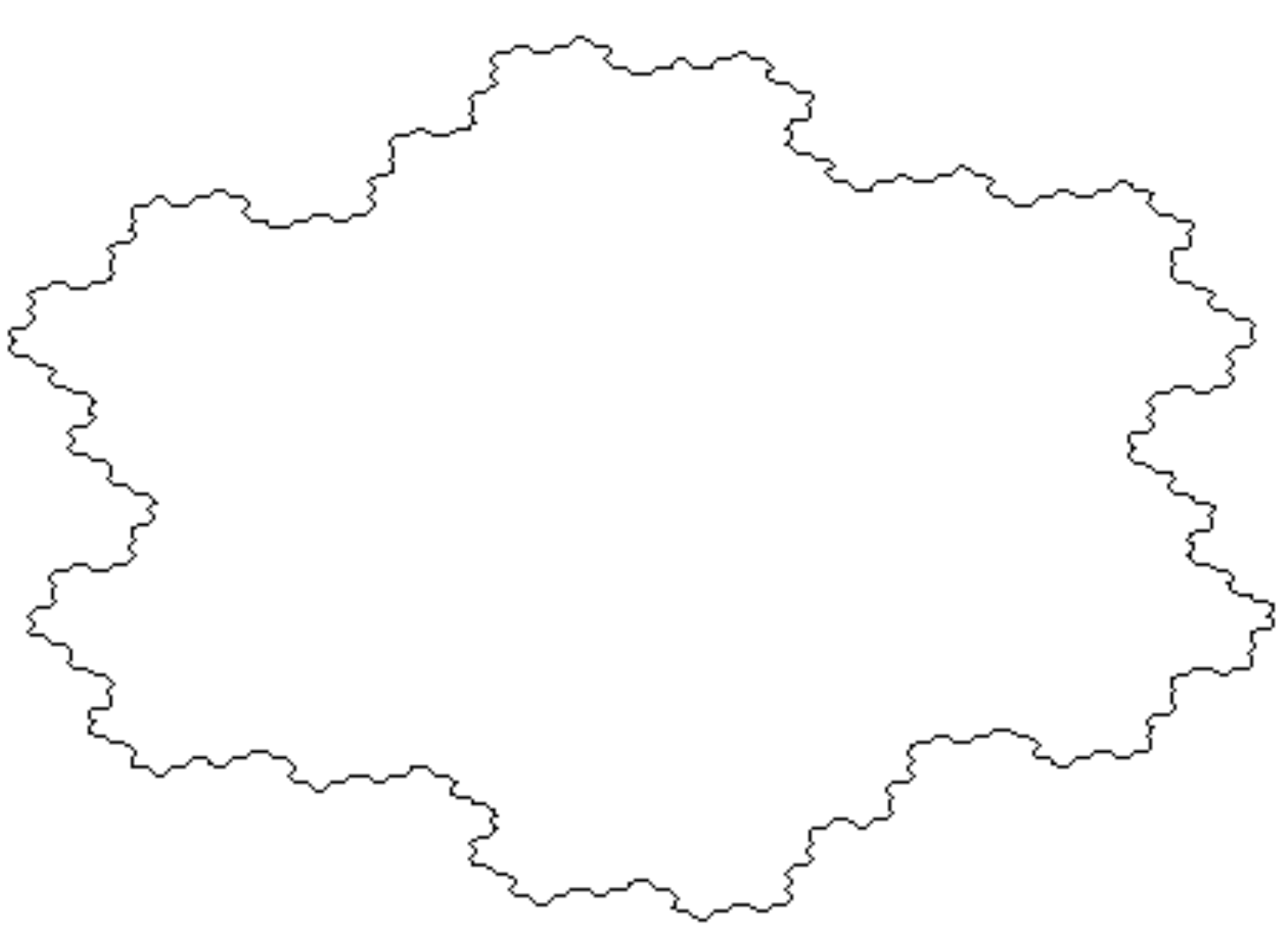}}
           			 \subfigure[]{\includegraphics[width=0.24\textwidth]{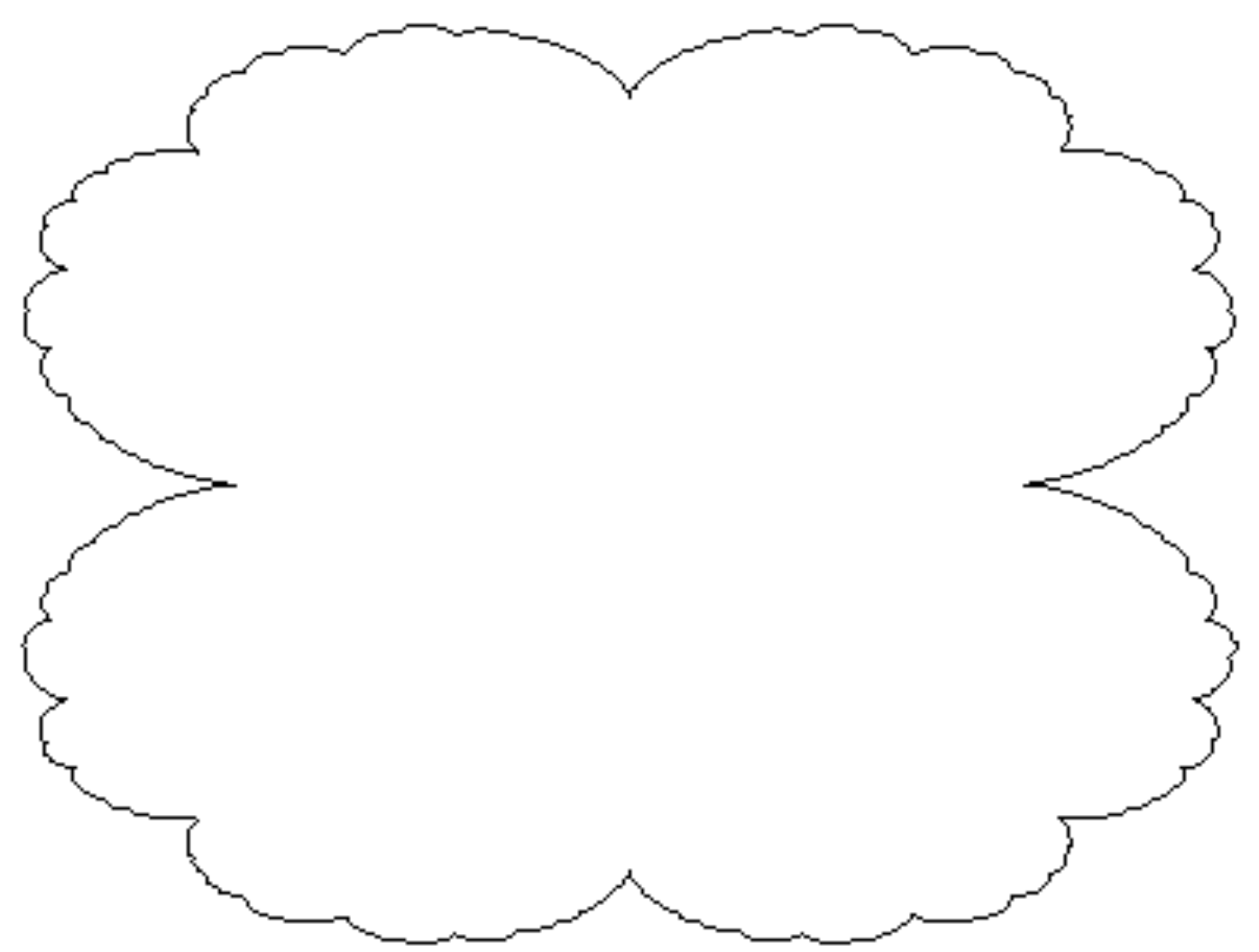}}}
           \mbox{\subfigure[]{\includegraphics[width=0.24\textwidth]{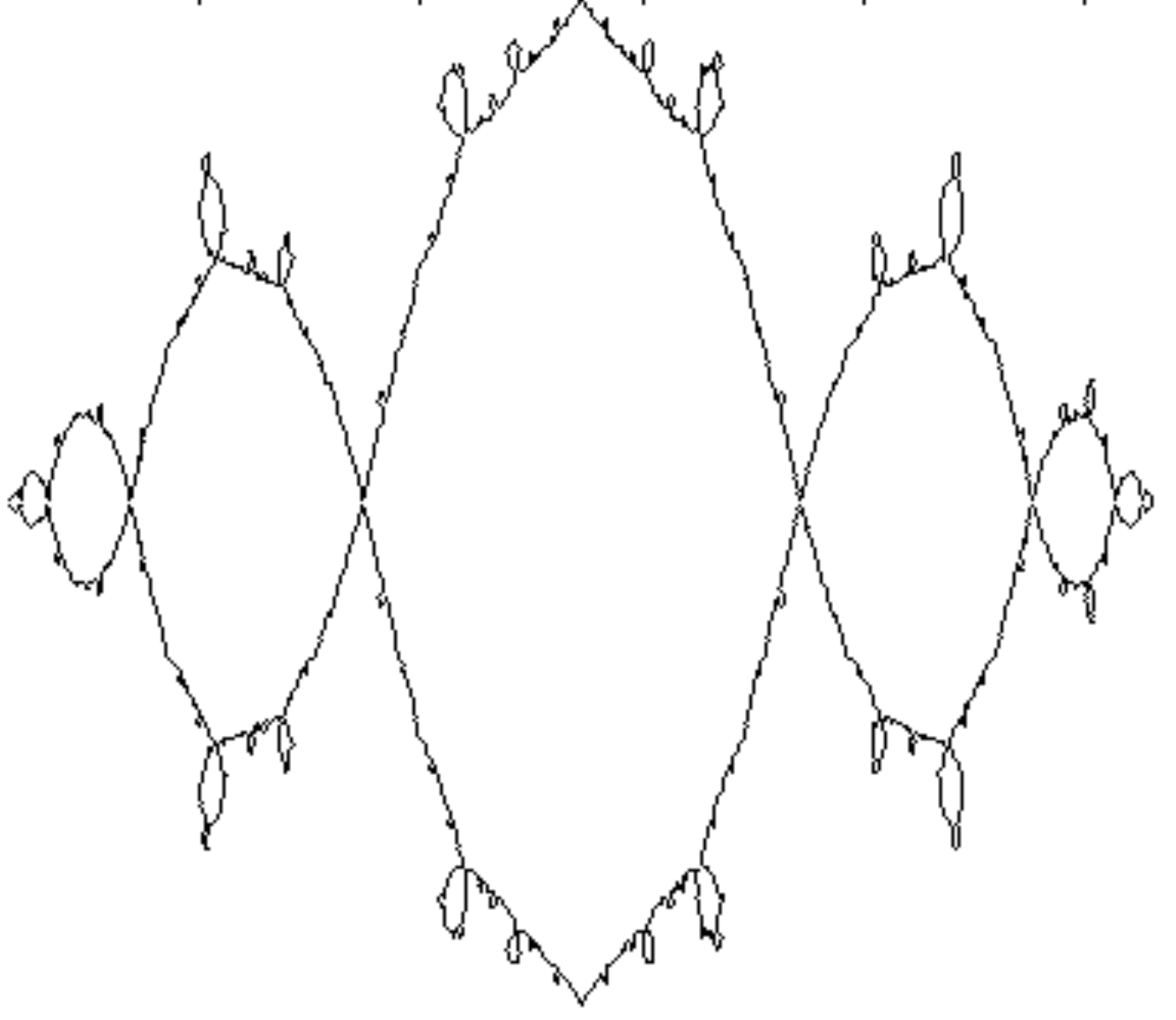}}
					       \subfigure[]{\includegraphics[width=0.24\textwidth]{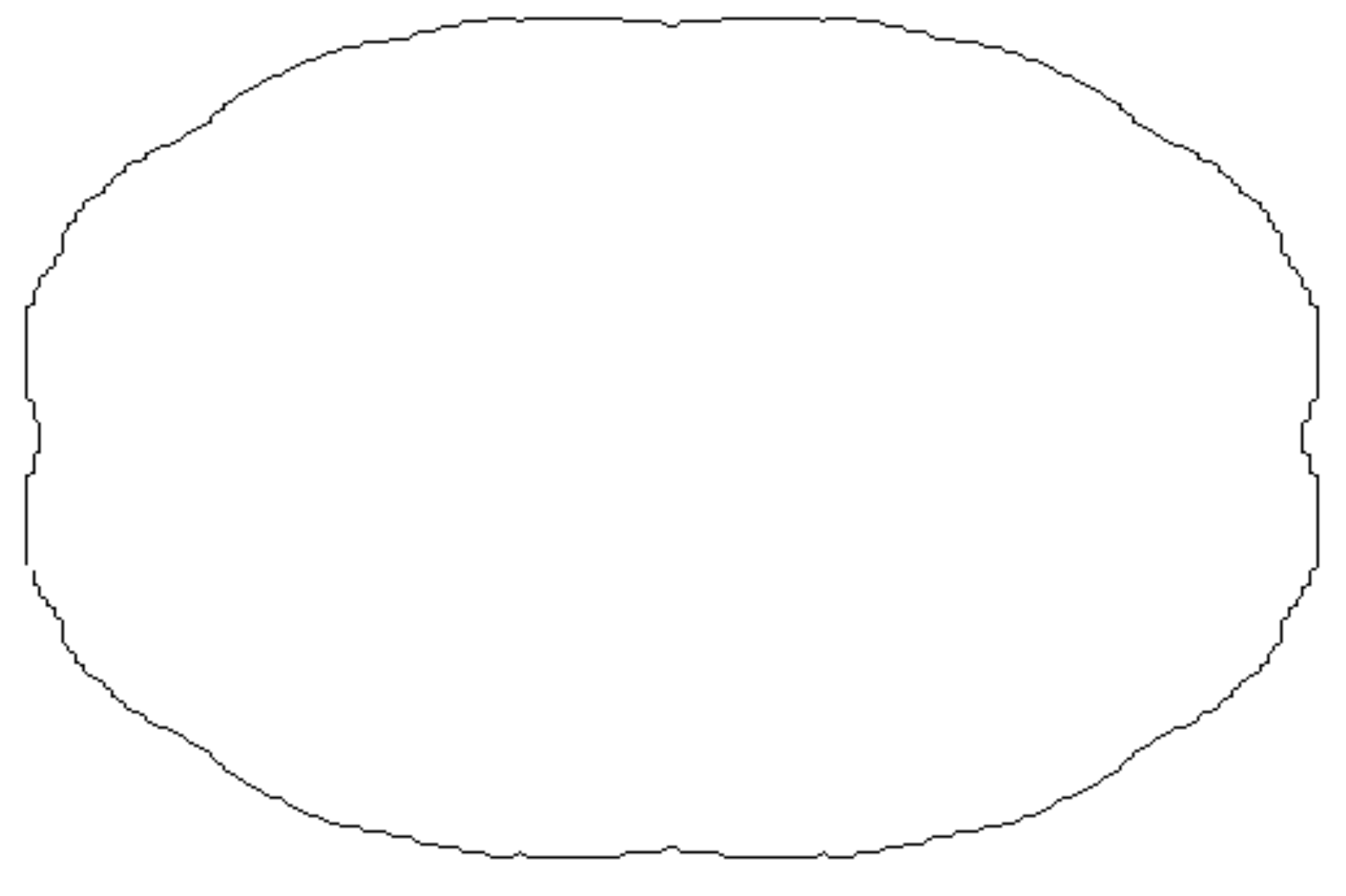}}
					       \subfigure[]{\includegraphics[width=0.24\textwidth]{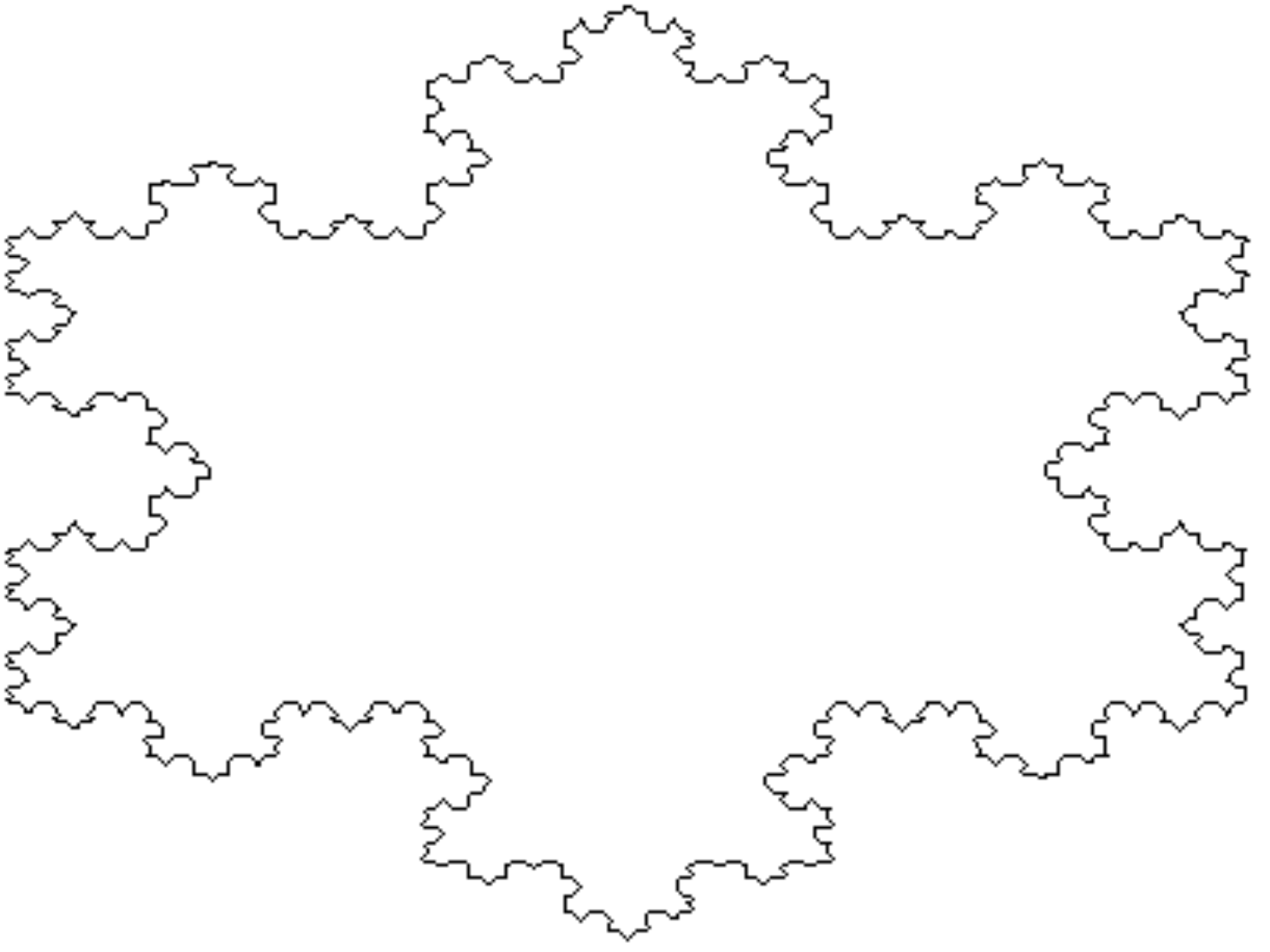}}
								 \subfigure[]{\includegraphics[width=0.24\textwidth]{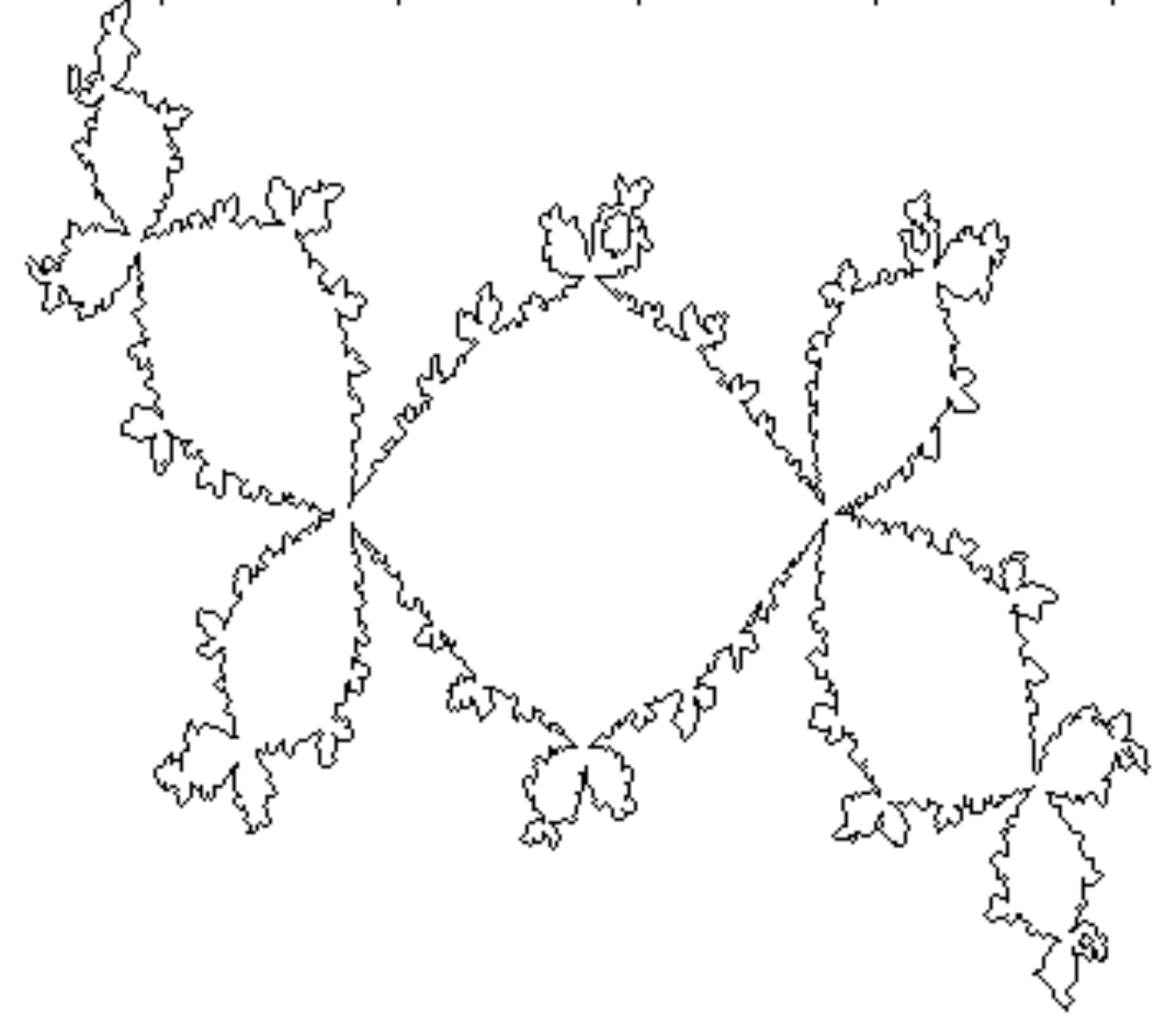}}}
					 \mbox{\subfigure[]{\includegraphics[width=0.24\textwidth]{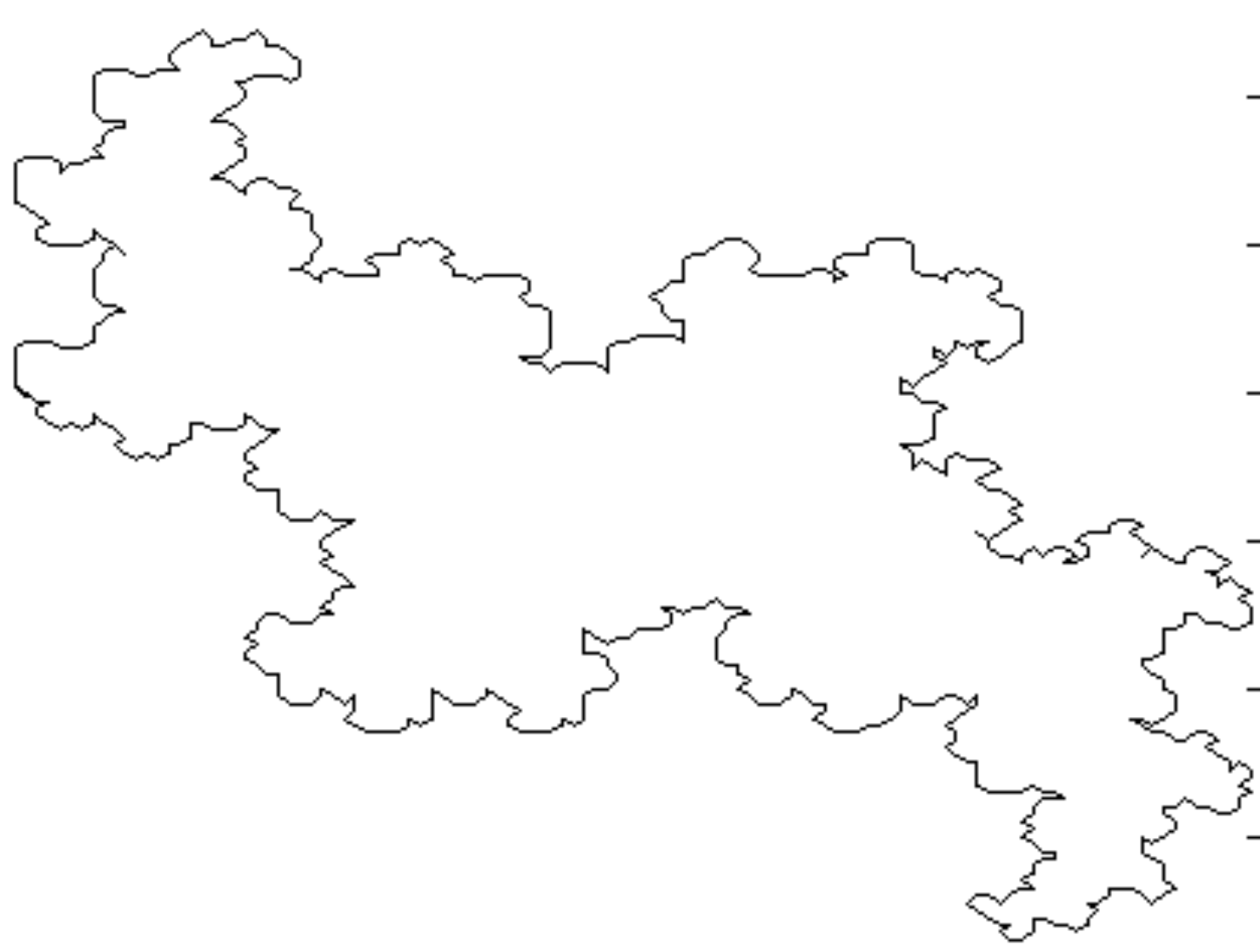}}
           			 \subfigure[]{\includegraphics[width=0.24\textwidth]{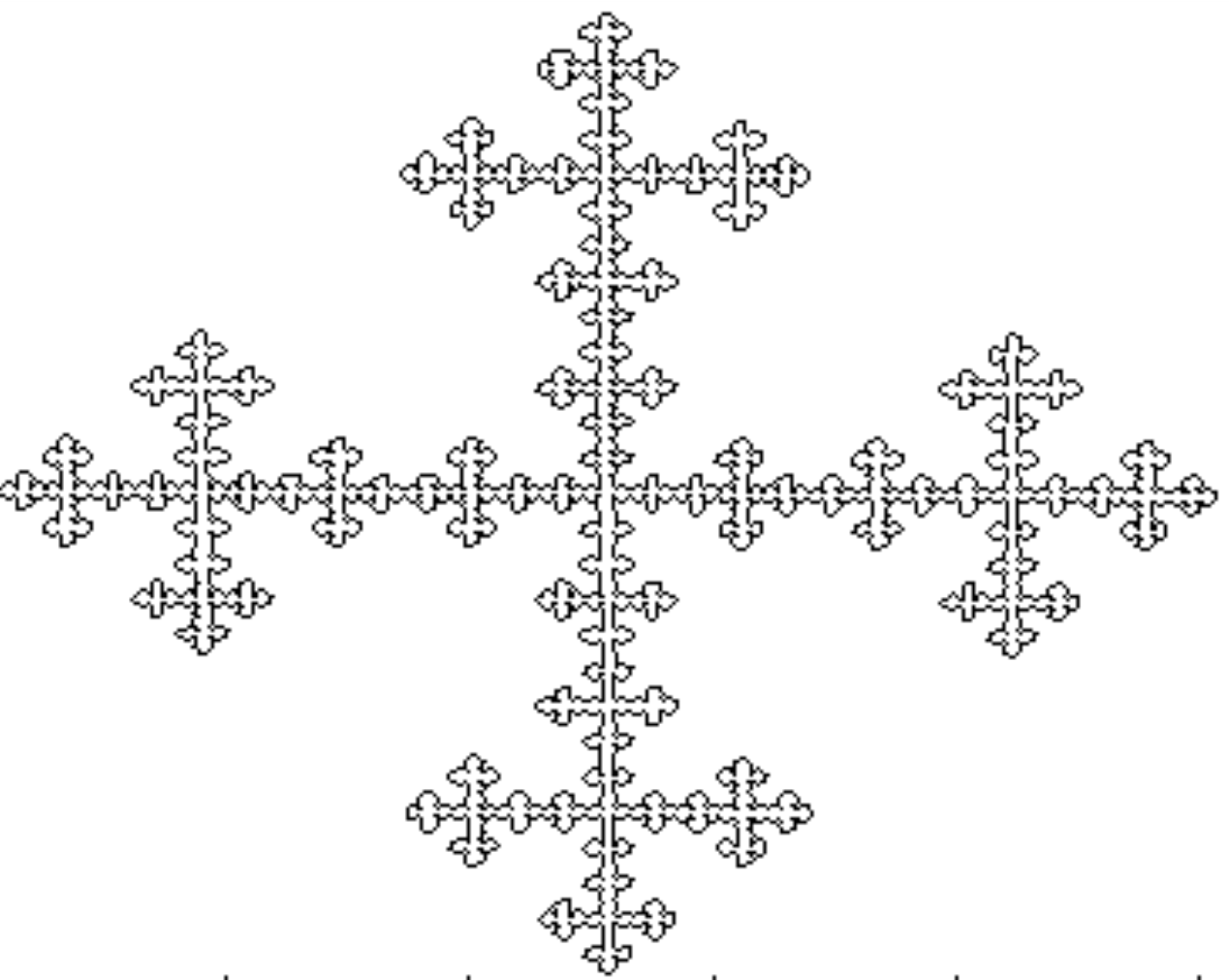}}}          			  				
           \caption{Fractal objects used in the experiments. a) Dragon curve. b) Fibonacci curve. c) Gosper curve. d) Julia set 1. e) Julia set 2. f) Julia set 3. g) Koch snowflake curve. h) Rabbit curve. i) Terdragon curve. j) Vicsek curve.}
           \label{fig:fractals}                                  
   \end{figure}

The fractals used in the experiments are examples of fractal closed curves, which can be represented by a parametric contour. The curves used are Dragon curve \cite{DK99} (Figure \ref{fig:fractals}(a)), Fibonacci curve \cite{M68} (Figure \ref{fig:fractals}(b)), Peano-Gosper curve \cite{M68} (Figure \ref{fig:fractals}(c)), the boundary of three variants of Julia sets \cite{D86} (Figure \ref{fig:fractals}(d), (e) and (f)), Koch snowflake curve \cite{LP04} (Figure \ref{fig:fractals}(g)), Douady-Rabbit curve \cite{D86} (Figure \ref{fig:fractals}(h)), Terdragon curve \cite{DK99} (Figure \ref{fig:fractals}(i)) and Vicsek curve \cite{ZZCYG08} (Figure \ref{fig:fractals}(j)).

The curves had its fractal dimension calculated by the method proposed here and the result was compared to the analytical Hausdorff dimension shown in the literature and to the fractal dimension calculated by other numerical methods well known, that is, Bouligand-Minkowski, box-counting and Fourier spectral method.

\begin{figure}[!htpb]
\centering
\includegraphics[width=0.7\textwidth]{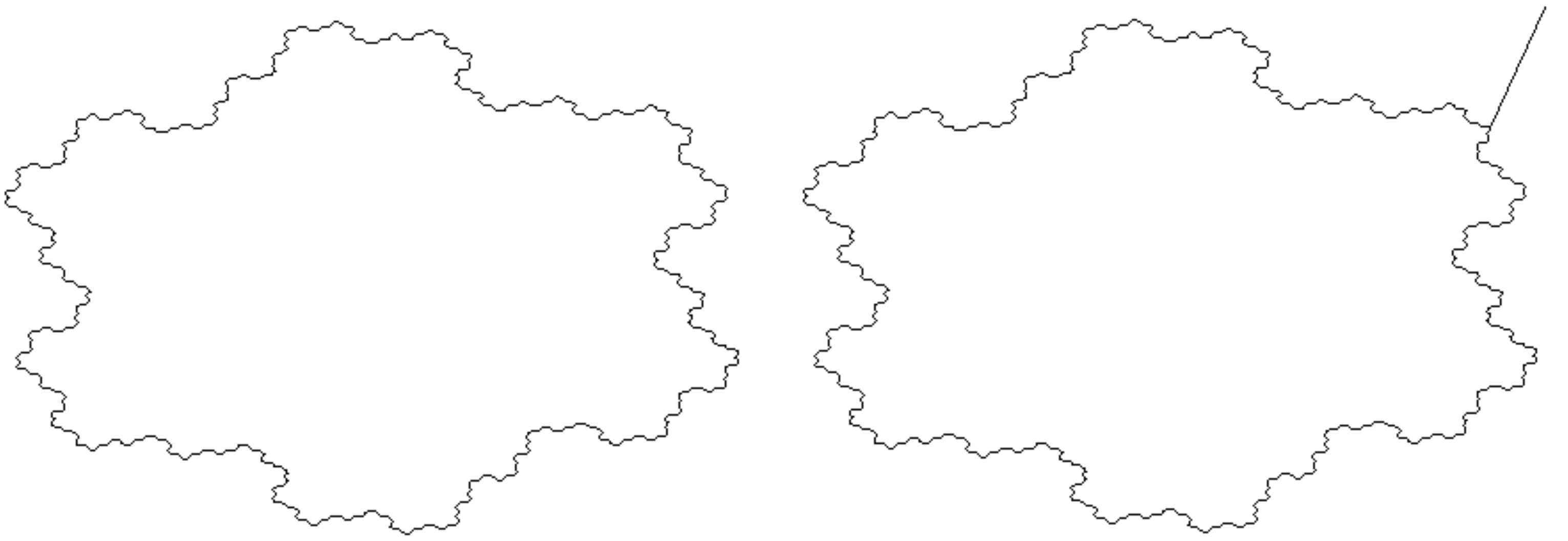}
\caption{To the left, an example of a fractal and to the right, the same fractal affected by the punctual random noise used in the experiments.}
\label{fig:noise}
\end{figure}

For a more robust test, the fractal curves were still submitted to rotation (by 45 and 90 degrees, representing a diagonal and a vertical perspective of the object), translation (by 10 and 20 pixels) and scale (by 0.5 and 2 times, representing a reduced and an extended copy of the original object). In a second moment, each fractal object was affected by a random punctual noise, as illustrated in the Figure \ref{fig:noise} and the fractal dimension was measured by the compared methods. Such random noises are commonly detected in objects represented in digital images due, for example, to a scale reduction. It is difficult to remove them from fractal objects once the noise reduction methods use to smooth the curve, reducing in this way the details which are fundamental in the representation of fractals. Thus, for practical purposes, it is important for the fractal dimension calculus method to be robust to such noise.

The accuracy of each fractal dimension method is measured by the average modular error of each calculated value relative to the analytical Hausdorff dimension value.

\section{Results}

The following tables present the fractal dimension calculated for the fractals affected by geometric transformations and noise. We can notice that the proposed technique is the unique totally invariant to the geometric transformations employed. Such fact is due to the intrinsic geometric invariance of the Fourier transform and to the manipulation of real values, avoiding rounding errors.

The Table \ref{tab:original} shows the fractal dimension calculated by each compared method and the Hausdorff dimension. The tables \ref{tab:rot45} and \ref{tab:rot90} show, respectively, the fractal dimensions of the fractal objects rotated by 45 and 90 degrees. The tables \ref{tab:scal05} and \ref{tab:scal2} show, respectively, the fractal dimension of the fractal objects scaled by a 0.5 and 2 factor. The tables \ref{tab:trans10} and \ref{tab:trans20} show the fractal dimension of the fractal objects translated, respectively, by 10 and 20 pixels. 

The Table \ref{tab:error} shows the average error (mean of the differences between the calculated and the waited value) of the fractal dimensions calculated by each compared method. The Table \ref{tab:noise} present the fractal dimension calculated for the fractals affected by the punctual noise. The Table \ref{tab:error_noise} shows the average error of the fractal dimensions calculated by the methods.

\begin{table*}[!htpb]
	\centering
	\scriptsize
	\caption{Fractal dimension of original fractals for each compared method. The underlined values are that which presented the best approximation to the theoretical Hausdorff dimension.}
		\begin{tabular}{c c c c c c}
			\hline
                 Fractal & Box-counting & Minkowski & Fourier & Proposed & Hausdorff\\
                 \hline
Dragon &1.4414 &1.4524 &1.3689 &\underline{1.5760} &1.5236\\
Fibonacci &1.3044    &1.3453    &1.2669    & \underline{1.2499} & 1.2465\\
Gosper &\underline{1.1284} &1.1217 &1.0685 &1.1213 &1.1292\\
Julia1 &\underline{1.0905} &1.0538 &1.1283 &1.1485 &1.0812\\
Julia2 &1.1465 &1.1972 &1.2146 &\underline{1.2509} &1.2683\\
Julia 3 &0.9677    &1.0166    &1.0680    &\underline{1.0076} & 1.0043\\
Koch &1.2479 &1.2497 &1.1080 &\underline{1.2722} &1.2619\\
Rabbit &1.2724 &\underline{1.3643} &1.2197 &1.3586 &1.3934\\
Terdragon &1.1871 &1.0949 &1.1162 &\underline{1.2313} &1.2619\\
Vicsek &1.3757 &\underline{1.4356} &1.3308 &1.5436 &1.4650\\
			\hline			
		\end{tabular}
	\label{tab:original}
\end{table*}

\begin{table*}[!htpb]
	\centering
	\scriptsize
	\caption{Fractal dimension of fractals rotated by 45 degrees.}
		\begin{tabular}{c c c c c c}
			\hline
                 Fractal & Box-counting & Minkowski & Fourier & Proposed & Hausdorff\\
                 \hline
Dragon &1.4339 &1.4535 &1.3839 &1.5760 &1.5236\\
Fibonacci &1.2525    &1.3479    &1.1742    &1.2499&1.2465\\
Gosper &1.1389 &1.1240 &1.0627 &1.1213 &1.1292\\
Julia1 &1.1140 &1.0566 &0.8218 &1.1485 &1.0812\\
Julia2 &1.1997 &1.1992 &0.9498 &1.2509 &1.2683\\
Julia 3 &1.0959    &1.0189    &0.9748    &1.0076&1.0043\\
Koch &1.2176 &1.2526 &1.1712 &1.2722 &1.2619\\
Rabbit &1.2783 &1.3657 &1.1106 &1.3586 &1.3934\\
Terdragon &1.2275 &1.0960 &0.9882 &1.2313 &1.2619\\
Vicsek &1.5059 &1.4368 &1.0311 &1.5436 &1.4650\\
			\hline			
		\end{tabular}
	\label{tab:rot45}
\end{table*}

\begin{table*}[!htpb]
	\centering
	\scriptsize
	\caption{Fractal dimension of fractals rotated by 90 degrees.}
		\begin{tabular}{c c c c c c}
			\hline
                 Fractal & Box-counting & Minkowski & Fourier & Proposed & Hausdorff\\
                 \hline
Dragon &1.4832 &1.4524 &1.4855 &1.5760 &1.5236\\
Fibonacci &1.3029    &1.3453    &1.2710    &1.2499&1.2465\\
Gosper &1.1428 &1.1217 &1.1224 &1.1213 &1.1292\\
Julia1 &1.0940 &1.0538 &1.1612 &1.1485 &1.0812\\
Julia2 &1.1468 &1.1972 &1.1226 &1.2509 &1.2683\\
Julia 3 &1.0913    &1.0166    &1.2325    &1.0076&1.0043\\
Koch &1.2487 &1.2497 &1.2141 &1.2722 &1.2619\\
Rabbit &1.3538 &1.3643 &1.2573 &1.3586 &1.3934\\
Terdragon &1.2410 &1.0949 &1.1132 &1.2313 &1.2619\\
Vicsek &1.4974 &1.4356 &1.3512 &1.5436 &1.4650\\
			\hline			
		\end{tabular}
	\label{tab:rot90}
\end{table*}

\begin{table*}[!htpb]
	\centering
	\scriptsize
	\caption{Fractal dimension of fractals scaled by 0.5 factor, that is, with halved size either in horizontal and vertical direction.}
		\begin{tabular}{c c c c c c}
			\hline
                 Fractal & Box-counting & Minkowski & Fourier & Proposed & Hausdorff\\
                 \hline
Dragon &1.3439 &1.5337 &1.2109 &1.5760 &1.5236\\
Fibonacci &1.2267    &1.2895    &1.1491    &1.2499&1.2465\\
Gosper &1.0669 &1.1078 &0.9326 &1.1213 &1.1292\\
Julia1 &1.0305 &1.0445 &1.0192 &1.1485 &1.0812\\
Julia2 &1.0946 &1.1760 &1.0957 &1.2509 &1.2683\\
Julia 3 &0.9157    &1.0133    &1.0706    &1.0076&1.0043\\
Koch &1.1760 &1.2279 &1.0777 &1.2722 &1.2619\\
Rabbit &1.2057 &1.3776 &1.0768 &1.3586 &1.3934\\
Terdragon &1.1455 &1.2571 &0.8698 &1.2313 &1.2619\\
Vicsek &1.3349 &1.4803 &1.1157 &1.5436 &1.4650\\
			\hline			
		\end{tabular}
	\label{tab:scal05}
\end{table*}

\begin{table*}[!htpb]
	\centering
	\scriptsize
	\caption{Fractal dimension of fractals scaled by a 2 factor.}
		\begin{tabular}{c c c c c c}
			\hline
                 Fractal & Box-counting & Minkowski & Fourier & Proposed & Hausdorff\\
                 \hline
Dragon &1.4948 &1.2672 &1.5031 &1.5760 &1.5236\\
Fibonacci &1.3586    &1.3354    &1.3830    &1.2499&1.2465\\
Gosper &1.1608 &1.1318 &1.1538 &1.1213 &1.1292\\
Julia1 &1.1218 &1.0711 &1.0345 &1.1485 &1.0812\\
Julia2 &1.1701 &1.2126 &1.1882 &1.2509 &1.2683\\
Julia 3 &1.0017    &0.9677    &1.0492    &1.0076&1.0043\\
Koch &1.2880 &1.2542 &1.0354 &1.2722 &1.2619\\
Rabbit &1.3042 &1.2495 &1.2831 &1.3586 &1.3934\\
Terdragon &1.2340 &0.8577 &1.3865 &1.2313 &1.2619\\
Vicsek &1.4046 &1.3318 &1.5783 &1.5436 &1.4650\\
			\hline			
		\end{tabular}
	\label{tab:scal2}
\end{table*}

\begin{table*}[!htpb]
	\centering
	\scriptsize
	\caption{Fractal dimension of fractals translated by 10 pixels from the original position either in horizontal and vertical direction.}
		\begin{tabular}{c c c c c c}
			\hline
                 Fractal & Box-counting & Minkowski & Fourier & Proposed & Hausdorff\\
                 \hline
Dragon &1.4396 &1.4524 &1.4137 &1.5760 &1.5236\\
Fibonacci &1.3045    &1.3453    &1.2832    &1.2499&1.2465\\
Gosper &1.1260 &1.1217 &1.1029 &1.1213 &1.1292\\
Julia1 &1.0861 &1.0538 &1.0868 &1.1485 &1.0812\\
Julia2 &1.1446 &1.1972 &1.1635 &1.2509 &1.2683\\
Julia 3 &0.9820    &1.0166    &1.0952    &1.0076&1.0043\\
Koch &1.2477 &1.2497 &1.0556 &1.2722 &1.2619\\
Rabbit &1.2723 &1.3643 &1.2078 &1.3586 &1.3934\\
Terdragon &1.1889 &1.0949 &1.0732 &1.2313 &1.2619\\
Vicsek &1.3767 &1.4356 &1.3319 &1.5436 &1.4650\\                
			\hline			
		\end{tabular}
	\label{tab:trans10}
\end{table*}

\begin{table*}[!htpb]
	\centering
	\scriptsize
	\caption{Fractal dimension of fractals translated by 20 pixels.}
		\begin{tabular}{c c c c c c}
			\hline
                 Fractal & Box-counting & Minkowski & Fourier & Proposed & Hausdorff\\
                 \hline
Dragon &1.4412 &1.4524 &1.4267 &1.5760 &1.5236\\
Fibonacci &1.3022    &1.3453    &1.2843    &1.2499&1.2465\\
Gosper &1.1297 &1.1217 &1.1158 &1.1213 &1.1292\\
Julia1 &1.0849 &1.0538 &1.0656 &1.1485 &1.0812\\
Julia2 &1.1457 &1.1972 &1.0821 &1.2509 &1.2683\\
Julia 3 &1.0025    &1.0166    &1.1686    &1.0076&1.0043\\
Koch &1.2425 &1.2497 &1.1116 &1.2722 &1.2619\\
Rabbit &1.2722 &1.3643 &1.2115 &1.3586 &1.3934\\
Terdragon &1.1882 &1.0949 &1.0503 &1.2313 &1.2619\\
Vicsek &1.3688 &1.4356 &1.3679 &1.5436 &1.4650\\                 
			\hline			
		\end{tabular}
	\label{tab:trans20}
\end{table*}

\begin{table*}[!htpb]
	\centering
	\scriptsize
	\caption{Average error in the calculus of each fractal dimension calculus method in the fractals affected by classical geometrical transforms.}
		\begin{tabular}{c c}
			\hline
                 Method              & Root Mean Square Error\\
                 \hline
								 Box-counting        & 0.2340\\
								 Bouligand-Minkowski & 0.2249\\
								 Classical Fourier   & 0.3610\\
								 Proposed method     & \underline{0.1269}\\
			\hline			
		\end{tabular}
	\label{tab:error}
\end{table*}

\begin{table*}[!htpb]
	\centering
	\scriptsize
	\caption{Calculus of fractal dimension in fractals affected by a puntual noise.}
		\begin{tabular}{c c c c c c}
			\hline
                 Fractal & Box-counting & Bouligand-Minkowski & Classical Fourier & Proposed method & Waited value\\                  \hline
Dragon &1.4461	&1.4623	&1.3712	&\underline{1.5404}	&1.5236\\
Fibonacci &1.3070    &1.3512    &1.2580    &\underline{1.2494} &1.2465\\
Gosper &1.1408	&1.1357	&1.1737	&\underline{1.1275}	&1.1292\\
Julia 1 &\underline{1.0965}	&1.0591	&1.1375	&1.1560	&1.0812\\
Julia 2 &1.1529	&1.1997	&1.2146	&\underline{1.2395}	&1.2683\\
Julia 3 &1.0494    &\underline{1.0315}    &1.1498    &1.1924 &1.0043\\
Koch &1.2503	&\underline{1.2542}	&1.1289	&1.2384	&1.2619\\
Rabbit &1.2737	&\underline{1.3642}	&1.2359	&1.3598	&1.3934\\
Terdragon &1.1871	&1.0949	&1.1162	&\underline{1.2313}	&1.2619\\
Vicsek &1.3767	&\underline{1.4358}	&1.3302	&1.5351	&1.465\\            
			\hline			
		\end{tabular}
	\label{tab:noise}
\end{table*}

\begin{table*}[!htpb]
	\centering
	\scriptsize
	\caption{Average error in the calculus of each fractal dimension method applied to fractals affected by noise.}	
		\begin{tabular}{c c}
			\hline
                 Method              & Root Mean Square Error\\
                 \hline
								 Box-counting        & 0.2307\\
								 Bouligand-Minkowski & 0.2244\\
								 Classical Fourier   & 0.3667\\
								 Proposed method     & \underline{0.2228}\\
			\hline			
		\end{tabular}
	\label{tab:error_noise}
\end{table*}

From the Table \ref{tab:original} we observe that the proposed method was the most precise. It approximated more faithfully the dimension of the most of the fractals analyzed. In its turn, the experiments involving classical geometric operations (rotation, translation and scale), depicted in tables from \ref{tab:rot45} to \ref{tab:trans20}, showed the robustness of the method. In the experiments, the fractal dimension calculated by the proposed technique was completely invariant to these operations. This invariance is easily explained by the intrinsic geometric invariance of the Fourier transform beyond the fact that the Fourier transform is an algebraic operation involving the manipulation of real values, which turns it unsusceptible to rounding errors as occurs with the other estimation methods, mainly that based on geometric analysis.

The Table \ref{tab:error} still shows that the proposed technique also is precise in the calculus of the fractal dimension, exhibiting the dimension value more similar to the theoretical value calculated by Hausdorff.

Finally, we observe that the tables \ref{tab:noise} and \ref{tab:error_noise} demonstrate the robustness to noise presented by the proposed method. This is explained by the fact that the Fourier transform concentrates the noises into the last coefficients, without a significant contribution to the dimension calculus, unlike geometrical methods which propagate the noise strongly.

\begin{table*}[!htpb]
	\centering
	\scriptsize
	\caption{Average computational time used by each method analyzed.}	
		\begin{tabular}{c c}
			\hline
                 Method              & Computational Time (s)\\
                 \hline
								 Box-counting        & 3.0394\\
								 Bouligand-Minkowski & 4.2518\\
								 Classical Fourier   & 4.2482\\
								 Proposed method     & \underline{0.4318}\\
			\hline			
		\end{tabular}
	\label{tab:time}
\end{table*}

Lastly, we verified the computational cost dispensed by each analyzed method, that, what the average time needed to calculate the fractal dimension of a fractal represented in a digital image by each estimation method. The result is shown in the Table \ref{tab:time}. The proposed method demonstrated a large advantage over the other methods being 7 times faster than Box-counting technique, the second faster algorithm.

\section{Conclusion}
This work proposed a novel method for the estimation of the fractal dimension of two-dimensional fractal (objects or curves) contours. The method proposed is based on the exponential relation between the power spectrum of the Fourier transform of a signal and the frequency variable. It is a method that can be easily implemented in a computer and presents a significant precision in the fractal dimension calculus as demonstrated by the results here exposed.

The results showed that the proposed technique presents a precision in the calculus of the dimension greater than traditional dimension estimation methods like Bouligand-Minkowski, box-counting and classical Fourier dimension. Besides, it was demonstrated the robustness of the method relative to the most common geometrical operation of rotation, translation and scale as relative to the presence of punctual noises.

The precision of the novel method is explained by the inherent precision of the Fourier transform and the robustness of the technique is interesting once it allows the estimation of the fractal dimension independently from the geometrical position of the object and even when the fractal is affected by noises, which is very common in the practice, either due to noise added in the process of capturing or digitalizing of the object. In this way, these good results suggest the use of the proposed method in applications which yield to the calculus of fractal dimension in varied fields of the science. For a practical approach, the method could be used to estimate the fractal dimension of signals or temporal series (in the real or complex domain), founding a wide range of applications in science and engineering.   

\section{Acknowledgements}
\label{sec:Acknowledgements}
Odemir M. Bruno gratefully acknowledges the financial support of CNPq (National Council for Scientific and Technological Development, Brazil) (Grant \#308449/2010-0 and \#473893/2010-0) and FAPESP (The State of S\~ao Paulo Research Foundation) (Grant \# 2011/01523-1). Jo\~ao B. Florindo is grateful to CNPq (National Council for Scientific and Technological Development, Brazil) for his doctorate grant.

\newpage


\end{document}